\documentclass[12pt]{article}
\usepackage{amsfonts,graphicx,amssymb,amsmath,amsthm,epsfig}
\usepackage{float}
\allowdisplaybreaks

\begin{document}
\title{\bf Analysis of Complexity Factor for Charged Dissipative Configuration
in Modified Gravity}
\author{M. Sharif$^1$ \thanks{msharif.math@pu.edu.pk} and K. Hassan$^2$ \thanks{komalhassan3@gmail.com}\\
$^1$ Department of Mathematics and Statistics, The University of Lahore,\\
1-KM Defence Road Lahore, Pakistan.\\
$^2$ Department of Mathematics, University of the Punjab,\\
Quaid-e-Azam Campus, Lahore-54590, Pakistan.}

\date{}

\maketitle
\begin{abstract}
In this paper, we determine the electromagnetic effects on the
complexity factor of radiating anisotropic cylindrical geometry in
the background of $f(G,\mathcal{T})$ theory. The self-gravitating
objects possessing inhomogeneous energy density, pressure
anisotropy, heat flux, charge and correction terms appear to
encounter the complexity producing phase. Herrera's orthogonal
splitting method is used to identify the scalar functions in which
the factor that incorporates all of the fundamental aspects of the
system is assumed to be the complexity factor. We also look at the
evolution of charged cylindrical matter source by selecting
homologous pattern as the most basic evolutionary mode. In addition
to this, homologous and complexity free conditions are utilized to
address dissipative as well as non-dissipative scenarios. The
complexity producing parameters throughout the evolutionary process
are assessed at the end. It is concluded that the complexity of the
astrophysical entities is elevated due to the contribution of charge
and modified terms of this theory.
\end{abstract}

\section{Introduction}

The universe is surrounded by a large number of physical entities,
which vary from unostentatious objects to gigantic and extravagant
astrophysical bodies. These cosmic entities provide the foundation
for cosmological analysis and support the expansion of the universe.
Hubble exploited a correlation between galaxy distance and recession
velocity and demonstrated that the universe is undergoing through
expansion phase. It is thought that the enigmatic substance known as
dark energy is accountable for the fast cosmic expansion. The rapid
evolution of the universe is indicated by a number of astronomical
events including supernovae and microwave background radiations
\cite{1a}. The foremost paradigm for understanding the current
accelerating expansion of the universe is described by the
$\Lambda$CDM model, however it has two major flaws: cosmic
coincidence and fine-tuning. In this regard, modified theories are
supposed to be an optimistic and promising approach in resolving
these problems and inspecting the role of dark energy in the
expanding cosmos.

The modified Gauss-Bonnet (GB) theory, commonly known as $f(G)$
gravity, was developed by Nojiri and Odintsov \cite{4} by
incorporating the generic GB function into the Einstein-Hilbert
action. The mathematical expression of the GB invariant is
$G=-4R^{\eta\vartheta}R_{\psi\chi}
+R^{\eta\vartheta\alpha\delta}R_{\eta\vartheta\alpha\delta}+R^2$.
Many cosmic and astrophysical phenomena that are taking place in the
universe are discussed in this theory. The scale factor in view of
two exponential functions is accommodated to investigate the
bouncing behavior and cosmic expansion through reconstructing the
$f(G)$ model by Bamba et al. \cite{5}. Abbas \cite{5a} looked into
the possibilities of developing cosmic objects in this theory and
inspected their physical attributes using the power-law model and
Krori-Barua ansatz. Sharif and Ramzan \cite{6b} discussed the
anisotropic stellar configurations and their remarkable physical
properties by assimilating the embedding class-1 method.

Sharif and Ikram \cite{7} added the trace of stress-energy tensor in
modified GB gravity action to develop a non-minimal gravity namely
$f(G,\mathcal{T})$ theory ($\mathcal{T}$ is the trace of
energy-momentum tensor). In the background of FRW cosmos, they
computed the energy constraints by reforming the de Sitter and
power-law models. The persistent characteristics of the Einstein
cosmos following homogeneous perturbations were explored by the same
authors \cite{7a}. According to Hossienkhani et al. \cite{7b},
anisotropy and weak energy constraint are  directly related to each
other in the anisotropic cosmos. With the help of $f(G,\mathcal{T})$
models, Shamir \cite{8a} evaluated the optimal bouncing solutions
corresponding to the selected equation of state.

When gravitational force is suppressed, the electromagnetic field is
crucial for understanding the stable and evolutionary mechanisms of
highly dense astronomical objects. A significant amount of charge is
required to defy gravity and retain the steady nature of compact
bodies. Ivanov \cite{9a} considered a charge configuration to devise
its solutions through regular (three classes) as well as general
methods. In order to figure out the influence of anisotropy on
charged distribution, Esculpi and Aloma \cite{9ab} related the
radial and tangential stresses in equation of state. Sharif and Naz
\cite{9b} examined the collapsing phenomenon of a charged
cylindrical distribution in $f(G)$ theory. We have utilized distinct
known isotropic solutions to illustrate the feasible features of
uncharged and charged anisotropic bodies via minimal geometric
deformation scheme, and then similar work is done for charged
distribution through extended geometric deformation technique in
$f(G,\mathcal{T})$ gravity \cite{9c}.

The intrinsic characteristics such as temperature, heat, energy
density, pressure, etc. of self-gravitating objects are mainly
accountable for the complex composition of celestial structures.
There is a need for the mathematical formula that takes into account
all significant physical variables in order to guage the complexity
of heavenly bodies. The notion of complexity in view of entropy and
information was presented by L{\'o}pez-Ruiz and his colleagues
\cite{10a}. This idea was first implemented on perfect crystal and
ideal gas. The arrangement of molecules in perfect crystal follows
the symmetric pattern and thus has no entropy, whereas the random
dispersion of particles in ideal gas results in maximum entropy. A
small region of perfect crystal is sufficient to elaborate on every
detail because the probability distribution does not contain any new
aspects. However, in ideal gas, the investigation of a tiny portion
leads to the maximum information (opposite to perfect crystal).
Thus, due to their opposing behavior, they are assigned zero
complexity. Based on how different probabilistic states diverge from
the system's equiprobable distribution, a novel definition of
complexity was constructed \cite{10b}. By following this idea, ideal
gas and perfect crystal were considered as complex free structures.

Later, the notion of complexity for static anisotropic geometrical
structure was modified by Herrera \cite{12} in order to encompass
all the crucial physical variables such as heat, pressure
anisotropy, energy density inhomogeneity and Tolman mass. The
Riemann tensor is divided using this method, resulting in the
production of several scalar functions in which the factor
accommodating each of the above-mentioned variables is assumed to be
the source of generating complexity. Sharif and Butt \cite{16}
studied the influence of charge on this factor in the case of
anisotropic sphere. Further, Herrera et al \cite{13} extended this
idea to non-static radiating source, in which the evolutionary
patterns have been an amazing increment. The same authors \cite{14}
also gauged the complexity factor for the axially symmetric
geometry. For the static cylindrical spacetime, Sharif and Butt
\cite{15} orthogonally decomposed the Riemann tensor to assess its
complexity.

The idea of studying complex structures has also gained attention in
modified theories. In this regard, Sharif et al. \cite{17a} employed
Herrera's strategy for the spherical symmetric structure and
ascertained the complexity factor in $f(G)$ theory. Yousaf et al.
\cite{17b} examined the influence of electromagnetic field on the
complexity factor of spherical distribution in the presence and
absence of charge under $f(G,\mathcal{T})$ gravity. We have also
decomposed the Riemann tensor orthogonally to compute
${\mathbb{Y}}_{\mathcal{T}F}$ (complexity producing source) in the
case of uncharged and charged cylindrical configuration. Moreover,
this technique has also been extended to dynamical spherical
(uncharged and charged) and cylindrical sectors \cite{17g}. There is
a significant collection of research on the evaluation of complexity
for various geometries in the context of several modified theories
\cite{18}.

This paper focuses on computing the complexity factor and evolution
modes of a dissipative cylindrical distribution influenced by
electromagnetic field and $f(G,\mathcal{T})$ corrections. The layout
of this paper is as follows. Section \textbf{2} provides the basic
characteristics of the matter source as well as the field equations
describing the system's dynamics. The structure scalars in terms of
physical variables are assessed through orthogonal breaking of the
Riemann tensor in section \textbf{3}. Two evolutionary patterns
(homologous and homogeneous) are discussed in section \textbf{4}.
Section \textbf{5} works for the evaluation of metric functions of
the considered geometry (for both dissipative and non-dissipative
scenarios). In section \textbf{6}, we talk about the reasons behind
the deviation of self-gravitating objects from complexity free
constraint. Finally, the main results are investigated in section
\textbf{7}.

\section{$f(G,\mathcal{T})$ Gravity and Physical Variables}

The Einstein-Maxwell action in the presence of Ricci scalar and
generalized $f(G,\mathcal{T})$ function is defined as
\begin{equation}\label{1}
\mathfrak{T}_{f(G,\mathcal{T})}=\frac{1}{16\pi}\int \sqrt{-g}
d^{4}x[f(G,\mathcal{T})+R]
+\int\sqrt{-g}(\mathcal{L}_\mathcal{M}+\mathcal{L}_\mathcal{E})d^{4}x,
\end{equation}
where $g$ demonstrates determinant of the metric tensor. The
quantities $\mathcal{L}_\mathcal{E}$ and $\mathcal{L}_\mathcal{M}$
represent the Lagrangian densities corresponding to the
electromagnetic field and ordinary matter, respectively. The
connection between the Lagrangian density and stress-energy tensor
is described through the relation
\begin{equation}\label{1a}
\mathcal{T}_{\eta\vartheta}=g_{\eta\vartheta}\mathcal{L}_\mathcal{M}-\frac{2\partial\mathcal{L}_\mathcal{M}}{\partial
g^{\eta\vartheta}}.
\end{equation}
Here, the modified field equations are computed as a result of
applying variational principle to \eqref{1} as
\begin{eqnarray}\nonumber
G_{\eta\vartheta}&=&8\pi
(\mathcal{T}_{\eta\vartheta}+\mathcal{S}_{\eta\vartheta})-(\Theta_{\eta\vartheta}+\mathcal{T}_{\eta\vartheta})f_{\mathcal{T}}(G,\mathcal{T})
+\frac{1}{2}g_{\eta\vartheta}f(G,\mathcal{T})+(4R_{\mu\vartheta}R^{\mu}_{\eta}\\\nonumber
&+&4R^{\mu\nu}R_{\eta\mu \vartheta \nu}-2RR_{\eta\vartheta}-2R^{\mu
\nu \gamma} _{\eta}R_{\vartheta \mu \nu
\gamma})f_{G}(G,\mathcal{T})+(4g_{\eta\vartheta}R^{\mu
\nu}\nabla_{\mu}\nabla_{\nu}\\\nonumber
&+&2R\nabla_{\eta}\nabla_{\vartheta}+4R_{\eta\vartheta}\nabla^{2}-2g_{\eta\vartheta}R\nabla^{2}
-4R^{\mu}_{\eta}\nabla_{\vartheta}\nabla_{\mu}
-4R^{\mu}_{\vartheta}\nabla_{\eta}\nabla_{\mu}\\\label{2}&-&4R_{\eta\mu
\vartheta \nu}\nabla^{\mu}\nabla^{\nu}) f_{G}(G,\mathcal{T}),
\end{eqnarray}
where
$\Theta_{\eta\vartheta}=-2\mathcal{T}_{\eta\vartheta}+{\mathcal{P}}g_{\eta\vartheta}$
and $\nabla^{2}=\Box=\nabla^{\mathfrak{l}}\nabla_{\mathfrak{l}}$
indicates the D' Alembert operator. The electromagnetic and Einstein
tensors are expressed as $\mathcal{S}_{\eta\vartheta}$ and
$G_{\eta\vartheta}$, respectively. Moreover, the partial
differentiation of an arbitrary function $f(G,\mathcal{T})$ is
denoted by $f_{G}=\frac{\partial f(G,\mathcal{T})}{\partial G}$ and
$f_{\mathcal{T}}=\frac{\partial f(G,\mathcal{T})}{\partial
\mathcal{T}}$.

The non-conservation of the energy-momentum tensor is caused by the
conjunction of geometrical expressions with the matter constituents.
This leads to the formation of an additional force that compels the
massive bodies in the gravitational field to trace the non-geodesic
path. Hence, we have the covariant differential of Eq.\eqref{2} as
\begin{eqnarray}\nonumber
\nabla^{\eta}\mathcal{T}_{\eta\vartheta}&=&\frac{f_{\mathcal{T}}(G,\mathcal{T})}{8\pi-f_{\mathcal{T}}(G,\mathcal{T})}
\bigg[-\frac{1}{2}g_{\eta\vartheta}\nabla^{\eta}\mathcal{T}
+\nabla^{\eta}\Theta_{\eta\vartheta}\\\label{2a}&+&(\Theta_{\eta\vartheta}+\mathcal{T}_{\eta\vartheta})\nabla^{\eta}(\ln
f_{\mathcal{T}}(G,\mathcal{T})) \bigg].
\end{eqnarray}
Alternatively, Eq.\eqref{2} can also be rewritten as
\begin{equation}\label{3a}
G_{\eta\vartheta}=8\pi
\mathcal{T}^{\textsf{(tot)}}_{\eta\vartheta}=8\pi(\mathcal{T^{(M)}}_{\eta\vartheta}+\mathcal{S}_{\eta\vartheta}
+\mathcal{T^{\textsf{(cor)}}}_{\eta\vartheta}),
\end{equation}
where the additional terms of $f(G,\mathcal{T})$ gravity indicated
by $\mathcal{T}^{\textsf{(cor)}}_{\eta\vartheta}$ have the following
form
\begin{eqnarray}\nonumber
\mathcal{T}^{\textsf{(cor)}}_{\eta\vartheta}&=&\frac{1}{8\pi}\left[\{(\mathcal{U}
+{\mathcal{P}}){\mathcal{V}}_{\eta}{\mathcal{V}}_{\vartheta}
+\Pi_{\eta\vartheta}+\varphi({\mathcal{V}}_{\eta}\chi_{\vartheta}
+\chi_{\eta}{\mathcal{V}}_{\vartheta})\}f_{\mathcal{T}}(G,\mathcal{T})
\right.\\\nonumber
&+&\left.(4R_{\mu\vartheta}R^{\mu}_{\eta}+4R^{\mu\nu}R_{\eta\mu
\vartheta \nu}-2RR_{\eta\vartheta}-2R^{\mu \nu \gamma}
_{\eta}R_{\vartheta \mu \nu
\gamma})f_{G}(G,\mathcal{T})\right.\\\nonumber
&+&\left.(4g_{\eta\vartheta}R^{\mu
\nu}\nabla_{\nu}\nabla_{\mu}-4R_{\eta\mu
\vartheta\nu}\nabla^{\nu}\nabla^{\mu}-4R^{\mu}_{\eta}\nabla_{\vartheta}\nabla_{\mu}
-2g_{\eta\vartheta}R\nabla^{2}\right.\\\label{4a}
&-&\left.4R^{\mu}_{\vartheta}\nabla_{\eta}\nabla_{\mu}+2R\nabla_{\eta}\nabla_{\vartheta}
+4R_{\eta\vartheta}\nabla^{2})f_{G}(G,\mathcal{T})\right]+\frac{g_{\eta\vartheta}f(G,\mathcal{T})}{2}.
\end{eqnarray}

The energy-momentum tensor for a normal matter distribution with
dissipation caused by heat flow is generally defined as
\begin{equation}\label{5c}
\mathcal{T^{(M)}}_{\eta\vartheta}=(\mathcal{U}+{{\mathcal{P}}}_{\bot})
{\mathcal{V}}_{\eta}{\mathcal{V}}_{\vartheta}+{{\mathcal{P}}}_{\bot}g_{\eta\vartheta}
+({{\mathcal{P}}}_{r}-{{\mathcal{P}}}_{\bot})\chi_{\eta}\chi_{\vartheta}
+{\varphi}_{\eta}{\mathcal{V}}_{\vartheta}+{\varphi}_{\vartheta}{\mathcal{V}}_{\eta},
\end{equation}
where the four vector along the radial direction $\chi^{\eta}$, four
velocity ${\mathcal{V}}^{\eta}$ and heat flux ${\varphi}^{\eta}$,
respectively, are interpreted as
\begin{equation}\label{5b}
\chi^{\eta}=\left(0,\textsf{Y}^{-1},0,0\right),\quad
{\mathcal{V}}^{\eta}=\left(\textsf{X}^{-1},0,0,0\right),\quad
{\varphi}^{\eta}=\left(0,{\varphi}\textsf{Y}^{-1},0,0\right),
\end{equation}
satisfying the relations $\chi^{\eta}\chi_{\eta}=1$,
$\chi^{\eta}{\mathcal{V}}_{\eta}=0$,
${\mathcal{V}}^{\eta}{\varphi}_{\eta}=0$ and
${\mathcal{V}}^{\eta}{\mathcal{V}}_{\eta}=-1$. Adding and
subtracting the term
$\frac{1}{3}{{\mathcal{P}}}_{r}({\mathcal{V}}_{\eta}{\mathcal{V}}_{\vartheta}+g_{\eta\vartheta})$
along with the relation ${\varphi}_{\iota}={\varphi}\chi_{\iota}$ in
Eq.\eqref{5c}, it follows that
\begin{align}\nonumber
\mathcal{T^{(M)}}_{\eta\vartheta}&=\mathcal{U}
{\mathcal{V}}_{\eta}{\mathcal{V}}_{\vartheta}
+\frac{1}{3}{{\mathcal{P}}}_{r}({\mathcal{V}}_{\eta}{\mathcal{V}}_{\vartheta}+g_{\eta\vartheta})
+\frac{2}{3}{{\mathcal{P}}}_{\bot}({\mathcal{V}}_{\eta}{\mathcal{V}}_{\vartheta}
+g_{\eta\vartheta})
-\frac{1}{3}{{\mathcal{P}}}_{r}({\mathcal{V}}_{\eta}{\mathcal{V}}_{\vartheta}+g_{\eta\vartheta})
\\\label{5e}&+\frac{1}{3}{{\mathcal{P}}}_{\bot}({\mathcal{V}}_{\eta}{\mathcal{V}}_{\vartheta}
+g_{\eta\vartheta})+({{\mathcal{P}}}_{r}-{{\mathcal{P}}}_{\bot})\chi_{\eta}\chi_{\vartheta}
+{\varphi}\chi_{\eta}{\mathcal{V}}_{\vartheta}+{\varphi}\chi_{\vartheta}{\mathcal{V}}_{\eta}.
\end{align}
Rearranging the above equation, we obtain
\begin{equation}\label{5a}
\mathcal{T^{(M)}}_{\eta\vartheta} =\mathcal{U}
{\mathcal{V}}_{\eta}{\mathcal{V}}_{\vartheta}+{{\mathcal{P}}}h_{\eta\vartheta}
+\Pi_{\eta\vartheta}+{\varphi}({\mathcal{V}}_{\eta}\chi_{\vartheta}+\chi_{\eta}{\mathcal{V}}_{\vartheta}),
\end{equation}
where
\begin{eqnarray}\label{101a}
\Pi_{\eta\vartheta}&=&\Pi\left(\chi_{\eta}\chi_{\vartheta}-\frac{h_{\eta\vartheta}}{3}\right),
\quad  \Pi={{\mathcal{P}}}_{r}-{{\mathcal{P}}}_{\bot},\\\label{101b}
{{\mathcal{P}}}&=&\frac{{{\mathcal{P}}}_{r}+2{{\mathcal{P}}}_{\bot}}{3},
\quad
h_{\eta\vartheta}={\mathcal{V}}_{\eta}{\mathcal{V}}_{\vartheta}+g_{\eta\vartheta}.
\end{eqnarray}
In the context of electromagnetic field, the stress-energy tensor is
defined as
\begin{equation}\label{3b}
\mathcal{S}_{\eta\vartheta}=\frac{1}{4\pi}\left(\mathbb{F}^{l}_{\eta}\mathbb{F}_{\vartheta
l}-\frac{1}{4}g_{\eta\vartheta}\mathbb{F}_{lm}\mathbb{F}^{lm}\right),
\end{equation}
where the Maxwell field tensor is indicated by
$\mathbb{F}_{\eta\vartheta}$ and the Maxwell field equations in
terms of tensorial form are portrayed as
\begin{equation}\nonumber
\mathbb{F}_{[\eta\vartheta;l]}=0,\quad
\mathbb{F}^{\eta\vartheta}_{~~;\vartheta}=4\pi \mathbb{J}^{\eta},
\end{equation}
where $\mathbb{J}^{\eta}$ shows the four current with
$\mathbb{J}^{\eta}= \texttt{e} {\mathcal{V}}^{\eta}$, and
$\texttt{e}$ implies the charge density.

The non-static cylindrical structure occupying anisotropy and heat
dissipation, enclosed by a hypersurface $\Sigma$ is delineated by
the line element
\begin{equation}\label{8a}
ds^{2}=-\textsf{X}^{2}dt^{2}+{\textsf{Y}}^{2}dr^{2}+{\textsf{Z}}^{2}d\theta^{2}+\alpha^2{\textsf{Z}}^2{dz^2},
\end{equation}
where the metric coefficients ${\textsf{X}},~{\textsf{Y}}$ and
${\textsf{Z}}$ depend upon $t$ and $r$, and $\alpha$ (a constant
term) has the dimension of inverse length. In the non-static
cylindrical system, the non-zero components of modified field
equations are
\begin{eqnarray}\label{9a}
8\pi({\textsf{X}}^{2}{\mathcal{U}}+\frac{\mathfrak{s}^2{\textsf{X}}^{2}}{8\pi
\textsf{Z}^4}+\mathcal{T}^{\textsf{(cor)}}_{00}) &=&
\frac{-{\textsf{X}}^2\left[\frac{2{\textsf{Z}}''}{{\textsf{Z}}}+\frac{{\textsf{Z}}'^2}{{\textsf{Z}}^2}
-\frac{2{\textsf{Z}}'{\textsf{Y}}'}{{\textsf{Z}}{\textsf{Y}}}\right]}{{\textsf{Y}}^2}
+\frac{\dot{{\textsf{Z}}}(\frac{2\dot{{\textsf{Y}}}}{{\textsf{Y}}}
+\frac{\dot{{\textsf{Z}}}}{{\textsf{Z}}})}{{\textsf{Z}}},\\\label{10a}
8\pi
(-{\varphi}{\textsf{X}}{\textsf{Y}}+\mathcal{T}^{\textsf{(cor)}}_{01})
&=& \frac{2{\textsf{X}}'\dot
{\textsf{Z}}}{{\textsf{X}}{\textsf{Z}}}+\frac{2{\textsf{Z}}'\dot
{\textsf{Y}}}{{\textsf{Z}}{\textsf{Y}}}-\frac{2\dot
{\textsf{Z}}'}{{\textsf{Z}}},
\\\nonumber 8\pi({\textsf{Y}}^{2}{{\mathcal{P}}}_{r}
-\frac{\mathfrak{s}^2{\textsf{Y}}^{2}}{8\pi
\textsf{Z}^4}+\mathcal{T}^{\textsf{(cor)}}_{11})&=&
\frac{-{\textsf{Y}}^2\left[-\frac{\dot{{\textsf{Z}}}}{{\textsf{Z}}}(\frac{2\dot{{\textsf{X}}}}{{\textsf{X}}}
-\frac{\dot{{\textsf{Z}}}}{{\textsf{Z}}})+\frac{2\ddot{{\textsf{Z}}}}{{\textsf{Z}}}\right]}{{\textsf{X}}^2}
+\bigg(\frac{2{\textsf{X}}'{\textsf{Z}}'}{{\textsf{X}}{\textsf{Z}}}+\frac{{\textsf{Z}}'^2}{{\textsf{Z}}^2}
\bigg),\\\label{11a}\\\nonumber
8\pi({\textsf{Z}}^{2}{{\mathcal{P}}}_\bot+\frac{\mathfrak{s}^2}{8\pi
\textsf{Z}^2}+\mathcal{T}^{\textsf{(cor)}}_{22})&=&
\frac{-{\textsf{Z}}^2}{{\textsf{A}
}^2}\left[\frac{\dot{{\textsf{Y}}}\dot{{\textsf{Z}}}}{{\textsf{Y}}{\textsf{Z}}}
+\frac{\ddot {\textsf{Y}}}{{\textsf{Y}}}+\frac{\ddot
{\textsf{Z}}}{{\textsf{Z}}}-\frac{\dot
{\textsf{X}}}{{\textsf{X}}}(\frac{\dot
{\textsf{Z}}}{{\textsf{Z}}}+\frac{\dot
{\textsf{Y}}}{{\textsf{Y}}})\right]\\\label{12a}
&+&\frac{{\textsf{Z}}^2}{{\textsf{Y}}^2}\left[(\frac{{\textsf{X}}'}{{\textsf{X}}}
-\frac{{\textsf{Y}}'}{{\textsf{Y}}})\frac{{\textsf{Z}}'}{{\textsf{Z}}}
-\frac{{\textsf{X}}'{\textsf{Y}}'}{{\textsf{X}}
{\textsf{Y}}}+\frac{{\textsf{X}}''}{{\textsf{X}}}+\frac{{\textsf{Z}}''}{{\textsf{Z}}}\right],
\end{eqnarray}
where . and $\prime$ stand for the partial derivatives with respect
to temporal and radial components, respectively. The additional
curvature terms, i.e.,
$\mathcal{T}^{\textsf{(cor)}}_{00},~\mathcal{T}^{\textsf{(cor)}}_{01},~\mathcal{T}^{\textsf{(cor)}}_{11}~$
and $\mathcal{T}^{\textsf{(cor)}}_{22}$ are given in
Eqs.(\ref{100})-(\ref{100c}) (Appendix \textbf{A}). In the present
system, the component of the Bianchi identities is obtained by using
Eq.\eqref{2a} as
\begin{align}\nonumber
\mathcal{T}^{\eta\vartheta}_{~~~;\vartheta}{\mathcal{V}}_{\eta}&=
\frac{-1}{{\textsf{X}}}\bigg\{\dot{\mathcal{U}}
+\frac{\dot
{\textsf{Y}}}{{\textsf{Y}}}\left({\mathcal{U}}+{{\mathcal{P}}}_{r}\right)+2\frac{\dot
{\textsf{Z}}}{{\textsf{Z}}}\left({\mathcal{U}}+{{\mathcal{P}}}_{\bot}\right)\bigg\}
\\\label{13a}&-\frac{1}{{\textsf{Y}}}\bigg\{{\varphi}'+2{\varphi}\bigg(\frac{\textsf{X}'}{\textsf{X}}
+\frac{\textsf{Z}'}{\textsf{Z}}\bigg) \bigg\}=\texttt{B}_{1}.
\end{align}

For the considered self-gravitating structure, the non-null
constituents of acceleration and shear tensor, respectively, are
\begin{equation}\label{15a}
a=\sqrt{a^{\eta}a_{\eta}}=\frac{{\textsf{X}}'}{{\textsf{Y}}{\textsf{X}}},
\quad a_{1}=\frac{{\textsf{X} }'}{{\textsf{X}}},
\end{equation}
\begin{equation}\label{18a}
\sigma_{11}=\frac{2}{3}{\textsf{Y}}^2\sigma, \quad
\sigma_{22}=\frac{\sigma_{33}}{\alpha^2}=-\frac{1}{3}{\textsf{Z}}^2\sigma,
\end{equation}
\begin{equation}\label{19a}
\sigma^{\eta\vartheta}\sigma_{\eta\vartheta}=\frac{2}{3}\sigma^2,~\sigma=\left(\frac{\dot
{\textsf{Y}}}{{\textsf{Y}}}-\frac{\dot
{\textsf{Z}}}{{\textsf{Z}}}\right)\frac{1}{{\textsf{X}}}.
\end{equation}
The expansion measures the rate of change in the volume with respect
to time, and its mathematical expression is \cite{51}
\begin{equation}\label{17a}
\Lambda_{\eta\vartheta}={\mathcal{V}}_{\alpha;\beta}h^{\alpha}_{(\eta}h^{\beta}_{\vartheta)},
\end{equation}
where the round bracket denotes the symmetric bracket in the indices
$\eta$ and $\vartheta$, while the trace of $\Lambda_{\eta\vartheta}$
is the expansion scalar \cite{13} and is given as
\begin{equation}\label{17b}
\Lambda={\mathcal{V}}^{\eta}_{~;\eta}=\left(2\frac{\dot
{\textsf{Z}}}{{\textsf{Z}}}+\frac{\dot
{\textsf{Y}}}{{\textsf{Y}}}\right)\frac{1}{{\textsf{X}}}.
\end{equation}
With the help of Eq.\eqref{10a}, one can obtain the following
expression as
\begin{equation}\label{21a}
4\pi\left({\varphi}{\textsf{Y}}-\frac{\mathcal{T}^{\textsf{(cor)}}_{01}}{{\textsf{X}}}\right)=-\sigma
\frac{{\textsf{Z}}'}{{\textsf{Z}}}+\frac{1}{3}\left(\Lambda-\sigma\right)'=\frac{{\textsf{Z}}'}{{\textsf{Y}}}
\left[\frac{1}{3}D_{{\textsf{Z}}}\left(\Lambda-\sigma\right)
-\frac{\sigma}{{\textsf{Z}}}\right],
\end{equation}
where
$D_{{\textsf{Z}}}=\frac{1}{{\textsf{Z}}'}\frac{\partial}{\partial
r}$ stands for the proper radial derivative. The mass of the
cylindrical structure is computed using the C-energy formula
proposed by Thorne \cite{19b} as
\begin{equation}\label{22a}
m=\frac{1}{2}{\textsf{Z}}^3R_{232}^{3}=\frac{\mathfrak{s}^2}{2\textsf{Z}}+\left[\frac{1}{4}
-\left(\frac{{\textsf{Z}}'}{{\textsf{Y}}}\right)^2+\left(\frac{\dot
{\textsf{Z}}}{{\textsf{X}}}\right)^2\right]\frac{{\textsf{Z}}}{2}.
\end{equation}

We propose the proper time derivative
($D_{T}=\frac{1}{{\textsf{X}}}\frac{\partial}{\partial t}$) to
understand the dynamical nature of charged cylinder. In the
collapsing phenomenon, the radius of an astronomical entity shrinks
when gravity dominates over outward pressure. As a result, the
interior fluid has negative velocity represented by
\begin{equation}\label{23a}
U=D_{T}{\textsf{Z}}< 0.
\end{equation}
The energy of the inner region is associated with the velocity by
the following relation
\begin{equation}\label{25a}
E\equiv\frac{{\textsf{Z}}'}{{\textsf{Y}}}=\left(\frac{1}{4}+U^2-\frac{2m}{{\textsf{Z}}}
+\frac{\mathfrak{s}^2}{\textsf{Z}^2}\right)^\frac{1}{2}.
\end{equation}
The energy distinction within the cylindrical object is estimated by
utilizing the time derivative ($D_{T}$). In view of time derivative,
the mass function \eqref{22a} along with the field equations
\eqref{10a} and \eqref{11a} takes the form
\begin{equation}\label{27a}
D_{T}m=\frac{\mathfrak{s}\dot{\mathfrak{s}}}{\textsf{X}\textsf{Z}}
-\frac{\mathfrak{s}^2\dot{\textsf{Z}}}{2\textsf{X}\textsf{Z}^2}-4\pi\left[\left({{\mathcal{P}}}_{r}-\frac{\mathfrak{s}^2}
{8\pi\textsf{Z}^4}+\frac{\mathcal{T}^{\textsf{(cor)}}_{11}}{{\textsf{Y}}^2}\right)U
+\left({\varphi}-\frac{\mathcal{T}^{\textsf{(cor)}}_{01}}{{\textsf{X}}{\textsf{Y}}}\right)E\right]
{\textsf{Z}}^2+\frac{\dot{\textsf{Z}}}{8\textsf{X}}.
\end{equation}
In terms of proper radial differentiation, we have used the field
equations \eqref{9a} and \eqref{10a}, turning the mass function as
\begin{equation}\label{28a}
D_{{\textsf{Z}}}m=\frac{\mathfrak{s}\mathfrak{s}'}{\textsf{Z}\textsf{Z}'}
-\frac{\mathfrak{s}^2}{2\textsf{Z}^2}+4\pi\left[\left({\mathcal{U}}+\frac{\mathfrak{s}^2}{8\pi\textsf{Z}^4}
+\frac{\mathcal{T}^{\textsf{(cor)}}_{00}}{{\textsf{X}}^2}\right)
+\left({\varphi}-\frac{\mathcal{T}^{\textsf{(cor)}}_{01}}{{\textsf{X}}{\textsf{Y}}}\right)\frac{U}{E}+\frac{1}{32\pi
{\textsf{Z}}^2}\right] {\textsf{Z}}^2.
\end{equation}
Consequently, from the above equation, we have
\begin{eqnarray}\nonumber
\frac{3m}{{\textsf{Z}}^3}&=&4\pi\left({\mathcal{U}}+\frac{\mathfrak{s}^2}{8\pi\textsf{Z}^4}
+\frac{\mathcal{T}^{\textsf{(cor)}}_{00}}{{\textsf{X}}^2}\right)
-\frac{4\pi}{{\textsf{Z}}^3}\int^{r}_{0}{\textsf{Z}}^3\left[D_{{\textsf{Z}}}\left({\mathcal{U}}
+\frac{\mathfrak{s}^2}{8\pi\textsf{Z}^4}
+\frac{\mathcal{T}^{\textsf{(cor)}}_{00}}{{\textsf{X}}^2}\right)
\right.\\\label{30a}
&-&\left.3\left({\varphi}-\frac{\mathcal{T}^{\textsf{(cor)}}_{01}}
{{\textsf{X}}{\textsf{Y}}}\right)
\frac{U}{{\textsf{Z}}E}\right]{\textsf{Z}}'dr+\frac{3\mathfrak{s}^2}
{2\textsf{Z}^4}+\frac{3}{8\textsf{Z}^2}.
\end{eqnarray}

The Weyl tensor is used to calculate the deformation that a
self-gravitating body would experience due to changes in the
gravitational field of a neighboring object. This tensor is
completely characterized by its two components, namely electric and
magnetic parts, as
\begin{equation*}
H_{\eta\vartheta}=\frac{1}{2}\eta_{\eta\nu\beta\gamma}C^{\beta\gamma}_{\vartheta\mu}\mathcal{V}^{\nu}\mathcal{V}^{\mu},
\quad
E_{\eta\vartheta}=C_{\eta\nu\vartheta\mu}\mathcal{V}^{\nu}\mathcal{V}^{\mu}.
\end{equation*}
For the symmetric cylindrical structure, the magnetic component
disappears while the electric part becomes
\begin{equation}\label{35a}
E_{\eta\vartheta}
=\epsilon(-\frac{h_{\eta\vartheta}}{3}+\chi_{\eta}\chi_{\vartheta}),
\end{equation}
where
\begin{eqnarray}\nonumber
\epsilon &=&\frac{1}{2{\textsf{X}}^2}\left[\frac{\ddot
{\textsf{Z}}}{{\textsf{Z}}}-\frac{\ddot
{\textsf{Y}}}{{\textsf{Y}}}-\left(\frac{\dot
{\textsf{Z}}}{{\textsf{Z}}}+\frac{\dot
{\textsf{X}}}{{\textsf{X}}}\right)\left(\frac{\dot
{\textsf{Z}}}{{\textsf{Z}}}-\frac{\dot
{\textsf{Y}}}{{\textsf{Y}}}\right)\right]-\frac{1}{2{\textsf{Z}}^2}+\frac{1}{2{\textsf{Y}}^2}
\left[-\frac{{\textsf{Z}}''}{{\textsf{Z}}}+\frac{{\textsf{X}}''}{{\textsf{X}}}
\right.\\\label{36a}
&+&\left.\left(\frac{{\textsf{Z}}'}{{\textsf{Z}}}
-\frac{{\textsf{X}}'}{{\textsf{X}}}\right)\left(\frac{{\textsf{Y}}'}{{\textsf{Y}}}
+\frac{{\textsf{Z}}'}{{\textsf{Z}}}\right)\right].
\end{eqnarray}
The contribution of tidal force to the charged source as a result of
$\epsilon$ is explained by
\begin{equation}\label{46a}
\frac{3m}{{\textsf{Z}}^3}=-\epsilon+\frac{3}{8\textsf{Z}^2}+\frac{3\mathfrak{s}^2}{2\textsf{Z}^4}
+4\pi\left[-\Pi^{\textsf{(tot)}}+\left({\mathcal{U}}+\frac{3\mathfrak{s}^2}{8\pi\textsf{Z}^4}
+\frac{\mathcal{T}^{\textsf{(cor)}}_{00}}{{\textsf{X}}^2}\right)
\right],
\end{equation}
where $\Pi^{\textsf{(tot)}}=\Pi+\Pi^{\textsf{(cor)}}$. It is crucial
to mention here that Eqs.\eqref{30a} and \eqref{46a} are expressed
with the same left hand value. The right side of equation
\eqref{30a} comprises the effects of inhomogeneous as well as
homogeneous energy density, charge and modified corrections, while
Eq.\eqref{46a} explains the role of anisotropic pressure and tidal
force in addition to energy density homogeneity, modified and
charged contributions. In the upcoming section, the term
$\frac{3m}{{\textsf{Z}}^3}$ will be found useful to express the
complexity generating factor in the required form.

\section{Orthogonal Splitting}

Bel \cite{22b} put forward the notion of orthogonal breaking of the
Riemann tensor. Herrera \cite{24a} used this approach to compute
several scalar functions displaying different physical aspects of
the system. These scalars disclose the complexity producing factors
and are called structure scalars. The conjunction of the Weyl
tensor, Ricci tensor and Ricci scalar in terms of the Riemann tensor
is delineated as
\begin{equation}\label{90a}
R^{\rho}_{\eta\vartheta\mu}={C}^{\rho}_{\eta\vartheta\mu}
+\frac{1}{2}R_{\eta\mu}\delta^{\rho}_{\vartheta}
+\frac{1}{2}R^{\rho}_{\vartheta}g_{\eta\mu}
-\frac{1}{2}R_{\eta\vartheta}\delta^{\rho}_{\mu}-\frac{1}{2}R^{\rho}_{\mu}g_{\eta\vartheta}
-\frac{1}{6}R\left(\delta^{\rho}_{\vartheta}g_{\eta\mu}\right.
-\left.g_{\eta\vartheta}\delta^{\rho}_{\mu}\right),
\end{equation}
and can be rewritten in view of matter distribution as
\begin{equation}\label{39a}
R^{\eta\gamma}_{\vartheta\delta}={C}^{\eta\gamma}_{\vartheta\delta}+16\pi
\mathcal{T}^{\textsf{(tot)}[\eta}_{[\vartheta}\delta^{\gamma]}_{\delta]}+8\pi
\mathcal{T}^{\textsf{(tot)}}\left(\frac{1}{3}\delta^{\eta}_{[\vartheta}\delta^{\gamma}_{\delta]}
-\delta^{[\eta}_{[\vartheta}\delta^{\gamma]}_{\delta]}\right).
\end{equation}
The following tensorial quantities are obtained by decomposing the
above equation as
\begin{eqnarray*}\nonumber
R^{\eta\gamma}_{(I)\vartheta\delta} &=&
16\pi\mathcal{U}\mathcal{V}^{[\eta}\mathcal{V}_{[\vartheta}\delta^{\gamma]}_{\delta]}
+2\mathcal{U}\mathcal{V}^{[\eta}\mathcal{V}_{[\vartheta}\delta^{\gamma]}_{\delta]}
+16\pi
{\mathcal{P}}h^{[\eta}_{[\vartheta}\delta^{\gamma]}_{\delta]}\\\label{70a}&+&2{\mathcal{P}}
\mathcal{V}^{[\eta}
\mathcal{V}_{[\vartheta}\delta^{\gamma]}_{\delta]}+8\pi(-\mathcal{U}+3{\mathcal{P}})
\left(\frac{1}{3}\delta^{\eta}_{[\vartheta}\delta^{\gamma}_{\delta]}-
\delta^{[\eta}_{[\vartheta}\delta^{\gamma]}_{\delta]}\right),
\\\nonumber
R^{\eta\gamma}_{(II)\vartheta\delta} &=&
16\pi\Pi^{[\eta}_{[\vartheta}\delta^{\gamma]}_{\delta]}+2\Pi^{[\eta}_{[\vartheta}\delta^{\gamma]}_{\delta]}+
\delta^{[\eta}_{[\vartheta}\delta^{\gamma]}_{\delta]}f\\\label{71a}&+&
8\delta^{[\eta}_{[\vartheta}\delta^{\gamma]}_{\delta]}
R^{ml}\nabla_{m}\nabla_{l}f_{G}-4R\delta^{[\eta}_{[\vartheta}\delta^{\gamma]}_{\delta]}\Box
f_{G},  \\\label{72a} R^{\eta\gamma}_{(III)\vartheta\delta} &=&
4\mathcal{V}^{[\eta}{\mathcal{V}_{[\vartheta}}E^{\gamma]}_{\delta]}
-\epsilon^{\eta\gamma}_{\beta}\epsilon_{\vartheta\delta\alpha}E^{\beta\alpha},\\\nonumber
R^{\eta\gamma}_{(IV)\vartheta\delta}
&=&2(R_{m\vartheta}R^{m\eta}\delta^{\gamma}_{\delta}-R_{m\delta}R^{m\eta}\delta^{\gamma}_{\vartheta}
+R_{m\vartheta}R^{m\gamma}
\delta^{\eta}_{\delta}+R_{m\delta}R^{m\gamma}\delta^{\eta}_{\vartheta})f_{G}\\\nonumber&+&2R^{ml}(R^{\eta}_{m\vartheta
l}\delta^{\gamma}_{\delta} -R^{\eta}_{m\delta
l}\delta^{\gamma}_{\vartheta}-R^{\gamma}_{m\vartheta
l}\delta^{\eta}_{\delta}+R^{\gamma}_{m\delta
l}\delta^{\eta}_{\vartheta})f_{G}-
R(R^{\eta}_{\vartheta}\delta^{\gamma}_{\delta}\\\nonumber&-&
R^{\eta}_{\delta}\delta^{\gamma}_{\vartheta}-R^{\gamma}_{\vartheta}\delta^{\eta}_{\gamma}
+R^{\gamma}_{\delta}\delta^{\eta}_{\vartheta})f_{G}-2(R^{\eta}_{m\vartheta
l}\delta^{\gamma}_{\delta}-R^{\eta}_{m\delta
l}\delta^{\gamma}_{\vartheta}-R^{\gamma}_{m\vartheta
l}\delta^{\eta}_{\delta}\\\nonumber&+&R^{\gamma}_{m\delta
l}\delta^{\eta}_{\vartheta})\nabla^{m}\nabla^{l}f_{G}+2(R^{\eta}_{\vartheta}\delta^{\gamma}_{\delta}
-R^{\eta}_{\delta}\delta^{\gamma}_{\vartheta}
-R^{\gamma}_{\vartheta}\delta^{\eta}_{\gamma}
+R^{\gamma}_{\delta}\delta^{\eta}_{\vartheta})\Box
f_{G}\\\nonumber&+&R(\delta^{\gamma}_{\delta}\nabla^{\eta}_{\vartheta}-\delta^{\gamma}_{\vartheta}\nabla^{\eta}_{\delta}
-\delta^{\eta}_{\delta}\nabla^{\gamma}_{\vartheta}+\delta^{\eta}_{\vartheta}\nabla^{\gamma}_{\delta})f_{G}
-2(R^{m\eta}\delta^{\gamma}_{\delta}
\nabla_{\vartheta}\nabla_{m}\\\nonumber&-&R^{m\eta}\delta^{\gamma}_{\vartheta}\nabla_{\delta}\nabla_{m}
-R^{m\gamma}\delta^{\eta}_{\delta}\nabla_{\vartheta}\nabla_{m}
+R^{m\gamma}\delta^{\eta}_{\vartheta}\nabla_{\delta}\nabla_{m})f_{G}
-2(R^{m}_{\vartheta}\\\nonumber&\times&\delta^{\gamma}_{\delta}\nabla^{\eta}\nabla_{m}
-R^{m}_{\delta}\delta^{\gamma}_{\vartheta}\nabla^{\eta}\nabla_{m}
-R^{m}_{\vartheta}\delta^{\eta}_{\delta}\nabla^{\gamma}\nabla_{m}
+R^{m}_{\delta}\delta^{\eta}_{\vartheta}\nabla^{\gamma}\nabla_{m})f_{G}
\\\nonumber&-&(R_{\vartheta lmn}R^{lmn\eta}\delta^{\gamma}_{\delta}-R_{\delta lmn}R^{lmn\eta}
\delta^{\gamma}_{\vartheta}-R_{\vartheta lmn}R^{lmn
\gamma}\delta^{\eta}_{\delta}+R_{\delta
lmn}\\\nonumber&\times&R^{lmn \gamma}\delta^{\eta}_{\vartheta})f_{G}
+\frac{1}{3}\left[(\mathcal{U}+{\mathcal{P}})f_{\mathcal{T}}+2f+4R_{m\alpha}R^{m\alpha}f_{G}
+4R^{lm}\right.\\\nonumber&\times&R^{\alpha}_{l\alpha m}f_{G}
-2R^{\beta}_{lmn}R^{lmn}_{\alpha}f_{G}-4R^{m\beta}\nabla_{m}\nabla_{\beta}f_{G}
-4R^{\alpha}_{l\alpha
m}\nabla^{m}\nabla^{l}f_{G}\\\nonumber&-&\left.2R\Box
f_{G}+16R^{lm}\nabla_{m}\nabla_{l}f_{G}
-4R^{m\alpha}\nabla_{\alpha}\nabla_{l}f_{G}
-2R^{2}f_{G}\right]\\\nonumber&-&\frac{\mathfrak{s}^2}{6r^4}h^{[\eta}_{[\vartheta}
\delta^{\gamma]}_{\delta]}
+\frac{\mathfrak{s}^2}{2r^4}\mathcal{V}^{[\eta}\mathcal
{V}_{[\vartheta}\delta^{\gamma]}_{\delta]}
-\frac{\mathfrak{s}^2}{r^4}\left(\chi^{[\eta}\chi_{[\vartheta}\delta^{\gamma]}_{\delta]}
-\frac{1}{3}h^{[\eta}_{[\vartheta}\delta^{\gamma]}_{\delta]}\right)
\\\label{150a}&+&\frac{\mathfrak{s}^2}{12\pi
r^4}\mathcal{V}^{[\eta}\mathcal{V}_{[\vartheta}\delta^{\gamma]}_{\delta]}
-\frac{\mathfrak{s}^2}{8\pi
r^4}\delta^{[\eta}_{[\vartheta}\delta^{\gamma]}_{\delta]},
\end{eqnarray*}
in which the following formulas are employed
\begin{equation}\nonumber
\epsilon_{\eta\gamma\vartheta}=\mathcal{V}^{\beta}\eta_{\beta\eta\gamma\vartheta},
\quad \epsilon_{\eta\gamma\vartheta}\mathcal{V}^{\vartheta}=0,
\end{equation}
\begin{equation}\nonumber
\epsilon^{\beta\gamma\alpha}\epsilon_{\alpha\eta\vartheta}=
\delta^{\beta}_{\eta}h^{\gamma}_{\vartheta}-\delta^{\beta}_{\eta}h^{\gamma}_{\vartheta}
+\mathcal{V}_{\eta}(\mathcal{V}^{\beta}\delta^{\gamma}_{\vartheta}-\delta^{\beta}_{\vartheta}\mathcal{V}^{\gamma}).
\end{equation}

Two tensors, i.e., $\mathbb{Y}_{\eta\vartheta}$ and
$\mathbb{X}_{\eta\vartheta}$ are introduced with the following form
\begin{eqnarray}\label{37a}
\mathbb{Y}_{\eta\vartheta}
&=&R_{\eta\gamma\vartheta\delta}{\mathcal{V}}^{\gamma}{\mathcal{V}}^{\delta},
\\\label{38a}\mathbb{X}_{\eta\vartheta}
&=&^{\ast}R^{\ast}_{\eta\gamma\vartheta\delta}{\mathcal{V}}^{\gamma}{\mathcal{V}}^{\delta}=
\frac{1}{2}\eta^{\epsilon\mu}_{\eta\gamma}
R^{\ast}_{\epsilon\mu\vartheta\delta}{\mathcal{V}}^{\gamma}{\mathcal{V}}^{\delta},
\end{eqnarray}
where $\eta^{\epsilon\mu}_{\eta\gamma}$ expresses the Levi-Civita
symbol and $R^{\ast}_{\eta\vartheta\gamma\delta}
=\frac{1}{2}\eta_{\epsilon\mu\gamma\delta}R^{\epsilon\mu}_{\eta\vartheta}$.
For the considered system, the tensors $\mathbb{Y}_{\eta\vartheta}$
and $\mathbb{X}_{\eta\vartheta}$ in view of matter sources are
computed as
\begin{eqnarray}\nonumber
\mathbb{Y}_{\eta\vartheta} &=&
\frac{1}{6}(\mathcal{U}+{\mathcal{P}})h_{\eta\vartheta}f_{\mathcal{T}}+\frac{\mathfrak{s}^2
h_{\eta\vartheta}}{3 r^4}+\frac{4\pi}{3}
(\mathcal{U}+3{\mathcal{P}})h_{\eta\vartheta}+E_{\eta\vartheta}\\\label{36a}&-&4\pi
\Pi_{\eta\vartheta}-\frac{\mathfrak{s}^2 }{
r^4}({\chi_{\eta}\chi_{\vartheta}-\frac{1}{3}h_{\eta\vartheta}})
+\frac{\Pi_{\eta\vartheta}}{2}f_{\mathcal{T}}
+\textsf{W}^{\textsf{(cor)}}_{\eta\vartheta},
\\\nonumber
\mathbb{X}_{\eta\vartheta}&=& \frac{8\pi \mathcal{U}
h_{\eta\vartheta}}{3}+\frac{\mathfrak{s}^2 h_{\eta\vartheta}}{3
r^4}+\frac{\Pi_{\eta\vartheta}}{2}f_{\mathcal{T}}-4\pi
\Pi_{\eta\vartheta}-\frac{\mathfrak{s}^2 }{
r^4}({\chi_{\eta}\chi_{\vartheta}-\frac{1}{3}h_{\eta\vartheta}})
\\\label{31}&-&E_{\eta\vartheta}
+\textsf{K}^{\textsf{(cor)}}_{\eta\vartheta}.
\end{eqnarray}
The modified terms $(\textsf{W}^{\textsf{(cor)}}_{\eta\vartheta}$
and $\textsf{K}^{\textsf{(cor)}}_{\eta\vartheta})$ are shown in
Appendix \textbf{B}. These tensors can be exhibited into their
scalar functions, i.e., ${\mathbb{Y}}_{\mathcal{T}},~
\mathbb{X}_{\mathcal{T}}$ (trace parts) and
${\mathbb{Y}}_{\mathcal{T}F},~\mathbb{X}_{\mathcal{T}F}$ (trace-free
parts) as
\begin{eqnarray}\label{40a}
\mathbb{Y}_{\eta\vartheta}
&=&\frac{h_{\eta\vartheta}{\mathbb{Y}}_{\mathcal{T}}}{3}+(\chi_{\eta}\chi_{\vartheta}
-\frac{h_{\eta\vartheta}}{3}){\mathbb{Y}}_{\mathcal{T}F},
\\\label{41a}
\mathbb{X}_{\eta\vartheta}
&=&\frac{h_{\eta\vartheta}{\mathbb{X}}_{\mathcal{T}}}{3}+(\chi_{\eta}\chi_{\vartheta}
-\frac{h_{\eta\vartheta}}{3}){\mathbb{X}}_{\mathcal{T}F}.
\end{eqnarray}

The structure scalars, for the system under consideration, are
evaluated as
\begin{eqnarray}\label{42a}
{\mathbb{Y}}_{\mathcal{T}}&=&
\frac{\mathfrak{s}^2}{\textsf{Z}^4}+4\pi\left({\mathcal{U}}+3{{\mathcal{P}}}_{r}-2\Pi
\right)+\frac{\left({\mathcal{U}}+{{\mathcal{P}}}\right)f_{\mathcal{T}}}{2}+\textsf{M}^{\textsf{(cor)}},
\\\label{43a} {\mathbb{Y}}_{\mathcal{T}F} &=& \epsilon-4\pi
\Pi-\frac{\mathfrak{s}^2}{\textsf{Z}^4}+\frac{\Pi}{2}f_{\mathcal{T}}+\textsf{L}^{\textsf{(cor)}},\\\label{44a}
{\mathbb{X}}_{\mathcal{T}}&=& 8\pi
({\mathcal{U}}+\frac{\mathfrak{s}^2}{8\pi\textsf{Z}^4})+\textsf{Q}^{\textsf{(cor)}},
\\\label{45a} {\mathbb{X}}_{\mathcal{T}F} &=&-\epsilon-4\pi \Pi-\frac{\mathfrak{s}^2}{\textsf{Z}^4}
+\frac{\Pi}{2}f_{\mathcal{T}},
\end{eqnarray}
where
$\textsf{L}^{\textsf{(cor)}}=\frac{\textsf{J}^{\textsf{(cor)}}_{\eta\vartheta}}{\chi_{\eta}\chi_{\vartheta}
-\frac{1}{3}h_{\eta\vartheta}}$. The quantities
$\textsf{J}^{\textsf{(cor)}}_{\eta\vartheta}$,$\textsf{M}^{\textsf{(cor)}}$,
and $\textsf{Q}^{\textsf{(cor)}}$ incorporate the additional terms
of this theory, exhibited in Appendix \textbf{B}. The influence of
electromagnetic field and energy density within the system is
described by ${\mathbb{X}}_{\mathcal{T}}$ while the contribution of
anisotropic stresses together with charge is allocated by the scalar
${\mathbb{Y}}_{\mathcal{T}}$. We can see that the scalar
${\mathbb{Y}}_{\mathcal{T}F}$ in Eq.\eqref{43a} only encompasses the
anisotropic pressure, so we need to rewrite this equation in a way
that this scalar contains the required parameters. For this purpose,
we substitute the value of $\frac{3m}{{\textsf{Z}}^3}$
(Eq.\eqref{30a}) in Eq.\eqref{46a} and after some calculations, the
scalar ${\mathbb{Y}}_{\mathcal{T}F}$ is obtained in the required
factors. Equations \eqref{36a} and \eqref{43a} express the scalar
${\mathbb{Y}}_{\mathcal{T}F}$ in the form of all fundamental
parameters of the structure through
\begin{eqnarray}\nonumber
{\mathbb{Y}}_{\mathcal{T}F}&=&\frac{4\pi}{{\textsf{Z}}^3} \int
{\textsf{Z}}^3\left[D_{{\textsf{Z}}}\left({\mathcal{U}}+\frac{\mathfrak{s}^2}{8\pi\textsf{Z}^4}
+\frac{\mathcal{T}^{\textsf{(cor)}}_{00}}{{\textsf{X}}^2}\right)
-3\left({\varphi}-\frac{\mathcal{T}^{\textsf{(cor)}}_{01}}{{\textsf{X}}{\textsf{Y}}}\right)
\frac{U}{{\textsf{Z}}E}\right]{\textsf{Z}}'dr
\\\label{47a}&-&4\pi\Pi-\frac{\mathfrak{s}^2}{2\textsf{Z}^4}+\frac{\Pi}{2}f_{\mathcal{T}}
-4\pi\Pi^{\textsf{(cor)}}-\textsf{L}^{\textsf{(cor)}}.
\end{eqnarray}
It can be observed from the above equation that this scalar is
associated with the energy density inhomogeneity, heat dissipation,
charge terms, anisotropic pressure and the extra curvature terms.
The impact of electromagnetic field, inhomogeneous energy density as
well as the modified terms can be seen in
${\mathbb{X}}_{\mathcal{T}F}$ as
\begin{equation}\label{48a}
{\mathbb{X}}_{\mathcal{T}F}=4\pi\Pi^{\textsf{(cor)}}-\frac{2\mathfrak{s}^2}{\textsf{Z}^4}
-\frac{4\pi}{{\textsf{Z}}^3}\int
{\textsf{Z}}^3\bigg[D_{{\textsf{Z}}}\bigg({\mathcal{U}}+\frac{\mathfrak{s}^2}{8\pi\textsf{Z}^4}
+\frac{\mathcal{T}^{\textsf{(cor)}}_{00}}{{\textsf{X}}^2}\bigg)
-3\bigg({\varphi}-\frac{\mathcal{T}^{\textsf{(cor)}}_{01}}{{\textsf{X}}{\textsf{Y}}}\bigg)
\frac{U}{{\textsf{Z}}E}\bigg]{\textsf{Z}}'dr.
\end{equation}
It has been noticed that the null ${\mathbb{Y}}_{\mathcal{T}F}$
factor together with different constraints (vanishing of radial and
tangential pressures, conformally flat condition, etc) has been
employed to compute the family of solutions corresponding to the
dynamical field equations \cite{13,50}. The,
${\mathbb{X}}_{\mathcal{T}F}$, can also be used as a restraint to
discuss the key features of the system accordingly.

\section{Different Modes of Evolution}

The behavior of the cosmological entities is significantly
influenced by substantial parameters, i.e., charge, pressure and
energy density. The existence of heat dissipation, pressure
anisotropy, additional terms and inhomogeneous energy density are
engrossed in ${\mathbb{Y}}_{\mathcal{T}F}$. We thus consider
${\mathbb{Y}}_{\mathcal{T}F}$ to be the complexity factor of the
radiating anisotropic distribution representing a non-static
geometry. The complexity free condition is obtained by substituting
${\mathbb{Y}}_{\mathcal{T}F}=0$ in order to produce the complex free
structure. The investigation of the modes of evolution is crucial
since the fluid under discussion is developing with time. In the
following sections, two evolutionary processes namely homologous
evolution and homogeneous expansion, are explored to investigate the
dynamical nature of self-gravitating bodies. The simplest pattern
that might help in the reduction of complexity during the
evolutionary process is also examined.

\subsection{Homologous Evolution}

The term homologous describes a phenomenon in which the entire
configuration has the similar pattern as a whole. When the matter
source of compact object falls internally at the center, it leads to
the collapsing phase. Homologous collapse is characterized by the
direct link between velocity and radial distance, which suggests
that all of the astrophysical entity's material sinks towards its
core at the same pace throughout the collapse. Hence, the structures
in which the collapse starts from the cores will produce fewer
gravitational radiations than homologously collapsing objects. An
alternative way of representing Eq.\eqref{21a} with the velocity of
the source is
\begin{equation}\label{52a}
D_{{\textsf{Z}}}\left(\frac{U}{{\textsf{Z}}}\right)=\frac{\sigma}{{\textsf{Z}}}+\frac{4\pi}{E}\left({\varphi}
-\frac{\mathcal{T}^{\textsf{(cor)}}_{01}}{{\textsf{X}}{\textsf{Y}}}\right),
\end{equation}
whose integration provides
\begin{equation}\label{53a}
U={\textsf{Z}}\int^{r}_{0}\left[\frac{\sigma}{{\textsf{Z}}}+\frac{4\pi}{E}\left({\varphi}
-\frac{\mathcal{T}^{\textsf{(cor)}}_{01}}{{\textsf{X}}{\textsf{Y}}}\right)
\right]{\textsf{Z}}'dr+\mathfrak{b(t)}{\textsf{Z}},
\end{equation}
where $\mathfrak{b(t)}$ is the integrating constant. After inserting
its value at the boundary, we have
\begin{equation}\label{54a}
U=\frac{U_{\Sigma}}{{\textsf{Z}}_{\Sigma}}{\textsf{Z}}-{\textsf{Z}}\int^{r_{\Sigma}}_{r}
\left[\frac{\sigma}{{\textsf{Z}}}+\frac{4\pi}{E}\left({\varphi}-\frac{\mathcal{T}^{\textsf{(cor)}}_{01}}{{\textsf{X}}{\textsf{Y}}}\right)
\right]{\textsf{Z}}'dr.
\end{equation}
It can be seen from the above equation that the heat flux and shear
scalar forced the matter distribution to deviate from the homologous
mode. The homologous condition ($U\sim {\textsf{Z}}$) \cite{22a} is
achieved if the integrand in Eq.\eqref{54a} vanishes which in turn
provides $U=\mathfrak{b(t)}{\textsf{Z}}$ and
$\mathfrak{b(t)}=\frac{U_{\Sigma}}{{\textsf{Z}}_{\Sigma}}$. The
condition of homologous evolution in $f(G,\mathcal{T})$ gravity is
written as
\begin{equation}\label{56a}
\frac{\sigma}{{\textsf{Z}}}+\frac{4\pi
{\textsf{Y}}}{{\textsf{Z}}'}\left({\varphi}-\frac{\mathcal{T}^{\textsf{(cor)}}_{01}}{{\textsf{X}}{\textsf{Y}}}\right)=0.
\end{equation}

\subsection{Homogeneous Expansion}

Here, we address another mode of evolution known as homogeneous
expansion and $\Lambda'=0$ is its required constraint. The
astrophysical objects undergoing expansion or contraction encounter
this phase when the corresponding collapsing or expanding rate does
not depend upon $r$. After combining this constraint with
Eq.\eqref{21a}, we have
\begin{equation}\label{57a}
4\pi\left({\varphi}-\frac{\mathcal{T}^{\textsf{(cor)}}_{01}}{{\textsf{X}}{\textsf{Y}}}\right)=-\frac{{\textsf{Z}}'}{{\textsf{Y}}}
\left[\frac{\sigma}{{\textsf{Z}}}+\frac{1}{3}D_{{\textsf{Z}}}(\sigma)\right].
\end{equation}
When we employ the homologous condition in the above equation, it
yields $D_{{\textsf{Z}}}(\sigma)=0$. At the core, the regularity
conditions provide $\sigma=0$, and its substitution in
Eq.\eqref{57a} gives rise to
\begin{equation}\label{59a}
{\varphi}=\frac{\mathcal{T}^{\textsf{(cor)}}_{01}}{{\textsf{X}}{\textsf{Y}}}.
\end{equation}
It displays the contribution of heat dissipation as a result of
$f(G,\mathcal{T})$ corrections. It can also be noticed that the
matter configuration evolving during homogeneous expansion was
shear-free and non-dissipative (in general relativity (GR)).

\section{Kinematical and Dynamical Considerations}

Now, in order to choose the most basic evolutionary mode, we examine
the behavior of several physical factors. For the sake of
simplicity, we use the assumption that ${\textsf{Z}}$ is taken as a
separable function of $r$ and $t$. The usage of homologous condition
\eqref{56a} together with Eq.\eqref{21a} provides
\begin{equation}\label{62a}
\left(\Lambda-\sigma\right)'=\left(\frac{3\dot
{\textsf{Z}}}{{\textsf{X}}{\textsf{Z}}}\right)'=0.
\end{equation}
One can notice from the above equation that it points towards the
geodesic nature of evolving fluid as ${\textsf{X}}'=0$, i.e., $a=0$.
Further, without any loss of generality, we can substitute
${\textsf{X}}=1$. On the contrary, the geodesic condition together
with the expressions of shear and expansion scalars give rise to
\begin{equation}\label{63a}
\Lambda-\sigma=\frac{3\dot {\textsf{Z}}}{{\textsf{Z}}}.
\end{equation}
The subsequent derivatives of Eq.\eqref{63a} lead to recover the
homologous condition. It means that cylindrical object evolves
homologously if and only if the matter source follows the geodesic
trajectory and vice versa. In GR, the absence of heat dissipation
(${\varphi}=0$) represents the matter without the contribution of
shear, whereas its impact in the current work is observed as
\begin{equation}\label{64a}
\sigma=4\pi\frac{\mathcal{T}^{\textsf{(cor)}}_{01}{\textsf{Z}}}{{\textsf{Z}}'}.
\end{equation}
We obtain $\mathcal{T}^{\textsf{(cor)}}_{01}=0$ when the matter
distribution undergoes homogeneous collapse in the absence of heat
flux and thus Eq.\eqref{57a} presents the shear scalar as
\begin{equation}\label{65a}
\sigma=\frac{12 \pi}{{\textsf{Z}}^3}\int
\frac{{\textsf{Z}}^3\mathcal{T}^{\textsf{(cor)}}_{01}}{{\textsf{X}}}dr
+\frac{\mathfrak{l(t)}}{{\textsf{Z}}^3},
\end{equation}
where the integration function is denoted by $\mathfrak{l(t)}$. In
other words, the homogeneous pattern corresponds to homologous
condition ($\sigma=0\Rightarrow U\sim {\textsf{Z}}$) if the
considered setup excludes the presence of heat dissipation and
correction terms. Hence, homologous evolution is regarded as the
most basic mode. As the system collapses, the mass and velocity that
are going through homologous mode are connected as
\begin{equation}\label{67a}
D_{T}U=\frac{\mathfrak{s}^2}{2\textsf{Z}^3}-\frac{m}{{\textsf{Z}}^2}
-4\pi\left({{\mathcal{P}}}_{r}
-\frac{\mathfrak{s}^2}{8\pi\textsf{Z}^4}
+\frac{\mathcal{T}^{\textsf{(cor)}}_{11}}{{\textsf{Y}}^2}\right)
{\textsf{Z}}+\frac{1}{8\textsf{Z}},
\end{equation}
while this equation can be rewritten by executing the value of
${\mathbb{Y}}_{TF}$ as
\begin{align}\nonumber
\frac{3D_{T}U}{{\textsf{Z}}}&={\mathbb{Y}}_{\mathcal{T}F}-\frac{\Pi}{2}f_{\mathcal{T}}
-\textsf{L}^{\textsf{(cor)}}+\frac{\mathfrak{s}^2}
{\textsf{Z}^4}+4\pi\Pi^{\textsf{(cor)}}-4\pi
\bigg[\bigg({\mathcal{U}}+\frac{\mathfrak{s}^2}{8\pi\textsf{Z}^4}
+\frac{\mathcal{T}^{\textsf{(cor)}}_{00}}{{\textsf{X}}^2}\bigg)
\\\label{68a}&+3\bigg({{\mathcal{P}}}_{r}-\frac{\mathfrak{s}^2}{8\pi\textsf{Z}^4}
+\frac{\mathcal{T}^{\textsf{(cor)}}_{11}}{{\textsf{Y}}^2}\bigg)-2\Pi\bigg].
\end{align}
After some manipulation, we can have the following result
\begin{equation}\label{71a}
\frac{\ddot {\textsf{Z}}}{{\textsf{Z}}}-\frac{\ddot
{\textsf{Y}}}{{\textsf{Y}}}={\mathbb{Y}}_{\mathcal{T}F}-\textsf{L}^{\textsf{(cor)}}+\frac{\mathfrak{s}^2}{\textsf{Z}^4}-\frac{\Pi}{2}
f_{\mathcal{T}}+4\pi\Pi^{\textsf{(cor)}}.
\end{equation}
One can observe that the charged cylindrical composition becomes
complex free if it satisfies
\begin{equation}\label{73a}
\frac{\ddot
{\textsf{Z}}}{{\textsf{Z}}}-4\pi\Pi^{\textsf{(cor)}}-\frac{\mathfrak{s}^2}{\textsf{Z}^4}-\frac{\ddot
{\textsf{Y}}}{{\textsf{Y}}}+\textsf{L}^{\textsf{(cor)}}+\frac{\Pi}{2}f_{\mathcal{T}}=0.
\end{equation}

Furthermore, we employ the homologous condition and zero complexity
constraint to compute the viable solutions in the presence and
absence of heat dissipation. The complications in the field
equations of $f(G,\mathcal{T})$ arise due to the inclusion of
non-linear terms and the considered system is non-static as well.
Thus, it is convenient to utilize the linear  $f(G,\mathcal{T})$
gravity model having the form $f(G,\mathcal{T})=\gamma G^n+\omega
\mathcal{T}$ \cite{23}. The parameters $\gamma$ and $\omega$ are the
real numbers along with $n>0$ and we have chosen $\gamma$ and $n$ to
be one. In the present configuration, three metric functions, i.e.,
${\textsf{X}}(t,r)$, ${\textsf{Y}}(t,r)$ and ${\textsf{Z}}(t,r)$
must be computed. However, the system constitutes two unknowns
(${\textsf{Y}}$, ${\textsf{Z}}$) due to the assumption
${\textsf{X}}=1$. This shows that there must be two equations to
calculate the unknowns. Hence, we are employing homologous and
vanishing complexity constraints to get the desired results. For the
non-dissipative configuration, the above-mentioned conditions are
evaluated as
\begin{eqnarray}\nonumber
&&\frac{1}{{{\textsf{Y}}^3 {\textsf{Z}}^2(\omega +8 \pi
)}}\left[\left(-2 {\textsf{Z}} \left(5 \ddot {\textsf{Y}} \dot
{\textsf{Z}}'+{\textsf{Z}}'\dot {\textsf{Y}} \right)+\dot
{\textsf{Y}} {\textsf{Z}}'\ddot {\textsf{Z}}\right){\textsf{Y}}+10
\dot {\textsf{Y}} \ddot {\textsf{Y}} {\textsf{Z}}
{\textsf{Z}}'\right.\\\label{77a}&&\left.+2 {\textsf{Y}}^2 \dot
{\textsf{Z}}' \left({\textsf{Z}}-4 \ddot
{\textsf{Z}}\right)\right]=0, \\\nonumber &&\frac{1}{2
{\textsf{Y}}^4 {\textsf{Z}}^4}\{2 {\textsf{Y}}^3\dot{{\textsf{Y}}}
{\textsf{Z}}^2 \dot{{\textsf{Z}}} \big(-4
\ddot{{\textsf{Z}}}+{\textsf{Z}}\big)+{\textsf{Y}} {\textsf{Z}}^2
{\textsf{Z}}^{'} \big(-2 {\textsf{Y}}^{'}
{\textsf{Z}}+\big(\dot{{\textsf{Y}}}^2+3 \ddot{{\textsf{Y}}}\big)
{\textsf{Z}}^{'}\\\nonumber&&+32 \dot{{\textsf{Y}}}
\dot{{\textsf{Z}}'}^2\big)-4 \dot{{\textsf{Y}}}^2 {\textsf{Z}}^2
{{\textsf{Z}}'}^2+{\textsf{Y}}^4 \big({\mathfrak{s}}^2-2
{\textsf{Z}}^2 \big(\dot{{\textsf{Z}}}^2+1\big)\big)+{\textsf{Y}}^2
{\textsf{Z}}^2 \big(2 {{\textsf{Z}}'}^2+16
\dot{{\textsf{Z}}'}^2\\\label{78a}&&-2 {\textsf{Z}}
\dot{{\textsf{Z}}}+{\mathfrak{s}}^2\big)\}=0.
\end{eqnarray}
The vanishing complexity condition for the second scenario, i.e.,
the dissipative source, is the same as it was for the
non-dissipative matter source, however, the homologous condition is
delineated as
\begin{eqnarray}\nonumber
{\varphi}&=&\frac{1}{16\pi^2 {\textsf{Z}}^4{\textsf{Y}}^5 (\omega
+1) }\big[{\textsf{Z}}'\big(-{\textsf{Y}} \dot {\textsf{Z}}+\dot
{\textsf{Y}}{\textsf{Z}}\big) \big({\textsf{Y}} \big({\textsf{Z}}
\big(40 \pi\dot {\textsf{Z}}' \ddot {\textsf{Y}}+\dot {\textsf{Y}}
{\textsf{Z}}' \big)-4 \pi\ddot {\textsf{Z}} \dot {\textsf{Y}}
{\textsf{Z}}'\big)\big.\big.\big.\\\label{79a}\big.\big.\big.&-&40
\pi \dot {\textsf{Y}} \ddot {\textsf{Y}} {\textsf{Z}}
{\textsf{Z}}'+{\textsf{Y}}^2 \big(-\dot {\textsf{Z}}'\big)
\big({\textsf{Z}}-32 \pi \ddot {\textsf{Z}}\big)\big)\big].
\end{eqnarray}

\section{Stability of Zero Complexity Condition}

The system may begin with null complexity and subsequently develop
uncertainties as a result of the existence of certain elements. By
using the approach described in \cite{24a}, we work on the evolution
equation in accordance with $\mathbb{Y}_{\mathcal{T}F}$ to assess
the intricacy producing factor with the help of Eqs.\eqref{13a},
\eqref{43a} and \eqref{45a} as
\begin{eqnarray}\nonumber
&&-4\pi\left({{\mathcal{P}}}_{r}+\frac{\mathcal{T}^{11\textsf{(cor)}}}
{{\textsf{Y}}^2}+{\mathcal{U}}+\mathcal{T}^{00\textsf{(cor)}}\right)\sigma
+\frac{\mathfrak{s}^2}{\textsf{Z}^4}+\frac{\mathfrak{s}^2\mathfrak{s}'}
{\textsf{Z}^4\textsf{Y}} -\frac{4\pi}{{\textsf{Y}}}\left[{\varphi}'-
(\frac{\mathcal{T}^{00\textsf{(cor)}}}{{\textsf{Y}}^2})'{\textsf{Y}}
\right.\\\nonumber&&\left.-({\varphi}
-\frac{\mathcal{T}^{01\textsf{(cor)}}}{{\textsf{Y}}^2})\frac{{\textsf{Z}}'}
{{\textsf{Z}}}\right] -\dot
{\mathbb{Y}}_{\mathcal{T}F}+\frac{3\mathfrak{s}\dot{\mathfrak{s}}}{\textsf{Z}^4}-\dot
{\textsf{L}}^{\textsf{(cor)}}-8\pi\dot
\Pi-4\pi\dot\Pi^{\textsf{(cor)}}-4\pi
\left(\mathcal{T}^{\textsf{(cor)}}_{00}\right)^.\\\nonumber&&-3\frac{\dot
{\textsf{Z}}}{{\textsf{Z}}}\left({\mathbb{Y}}_{\mathcal{T}F}-\textsf{L}^{\textsf{(cor)}}\right)-12\pi\big(
\mathcal{T}^{00\textsf{(cor)}}+\frac{\mathfrak{s}^2}{8\pi\textsf{Z}^4}\big)\frac{\dot
{\textsf{Z}}}{{\textsf{Z}}}-4\pi \texttt{B}_{1}+\frac{4\pi
\mathcal{T}^{00\textsf{(cor)}}{\textsf{Y}}'}{{\textsf{Y}}^2}\\\label{87a}&&+\left(12\pi
\mathcal{T}^{\textsf{(cor)}}_{00}-12\pi\Pi^{\textsf{(cor)}}+\frac{12\pi
\mathcal{T}^{\textsf{(cor)}}_{11}}{{\textsf{Y}}^2}\right)\frac{\dot
{\textsf{Z}}}{{\textsf{Z}}}-\frac{16\pi\Pi^{\textsf{(tot)}}\dot
{\textsf{Z}}}{{\textsf{Z}}}-\frac{5\mathfrak{s}^2\dot{\textsf{Z}}}{2\textsf{Z}^5}
=0.
\end{eqnarray}
By substituting
${\mathbb{Y}}_{\mathcal{T}F}=\sigma=\Pi^{\textsf{(tot)}}={\varphi}=0$
in Eq.\eqref{87a} at $t=0$ (initial time), the non-dissipative fluid
gives rise to
\begin{eqnarray*}\nonumber
&&-\dot {\mathbb{Y}}_{\mathcal{T}F}-4\pi
\left(\mathcal{T}^{\textsf{(cor)}}_{00}\right)^.-\dot
{\textsf{L}}^{\textsf{(cor)}}+4\pi(\frac{\mathcal{T}^{00\textsf{(cor)}}}{{\textsf{Y}}^2})'-4\pi\dot\Pi^{\textsf{(cor)}}-8\pi\dot
\Pi+3\textsf{L}^{\textsf{(cor)}}\\\nonumber&&\times\frac{\dot
{\textsf{Z}}}{{\textsf{Z}}}-4\pi \texttt{B}_{1}-12\pi\big(
\mathcal{T}^{00\textsf{(cor)}}+\frac{\mathfrak{s}^2}{8\pi\textsf{Z}^4}\big)\frac{\dot
{\textsf{Z}}}{{\textsf{Z}}}+\frac{4\pi
\mathcal{T}^{00\textsf{(cor)}}{\textsf{Y}}'}{{\textsf{Y}}^2}+\big(12\pi
\mathcal{T}^{\textsf{(cor)}}_{00}+\frac{12\pi
\mathcal{T}^{\textsf{(cor)}}_{11}}{{\textsf{Y}}^2}\big)\\&&\times\frac{\dot
{\textsf{Z}}}{{\textsf{Z}}}+\frac{\mathfrak{s}^2}{\textsf{Z}^4}+\frac{\mathfrak{s}^2\mathfrak{s}'}{\textsf{Z}^4\textsf{Y}}
+\frac{3\mathfrak{s}\dot{\mathfrak{s}}}{\textsf{Z}^4}-\frac{5\mathfrak{s}^2\dot{\textsf{Z}}}{2\textsf{Z}^5}=0.
\end{eqnarray*}
When the derivative of Eq.\eqref{47a} is used with the
aforementioned restriction at $t=0$, it results
\begin{eqnarray}\nonumber
&&\frac{4\pi}{{\textsf{Z}}^3}\int
^{r}_{0}[({\mathcal{U}}+\frac{\mathfrak{s}^2}{8\pi\textsf{Z}^4}+\frac{\mathcal{T}^{\textsf{(cor)}}_{00}}{{\textsf{X}}^2})^.]'=
4\pi(\frac{\mathcal{T}^{00\textsf{(cor)}}}{{\textsf{Y}}^2})'-4\pi
\left(\mathcal{T}^{\textsf{(cor)}}_{00}\right)^.+\frac{4\pi
\mathcal{T}^{00\textsf{(cor)}}{\textsf{Y}}'}{{\textsf{Y}}^2}\\\nonumber&&-12\pi\big(
\mathcal{T}^{00\textsf{(cor)}}+\frac{\mathfrak{s}^2}{8\pi\textsf{Z}^4}\big)\frac{\dot
{\textsf{Z}}}{{\textsf{Z}}}+3{\textsf{L}}^{\textsf{(cor)}}\frac{\dot
{\textsf{Z}}}{{\textsf{Z}}}-4\pi \texttt{B}_{1}+(12\pi
\mathcal{T}^{\textsf{(cor)}}_{00}+\frac{12\pi
\mathcal{T}^{\textsf{(cor)}}_{11}}{{\textsf{Y}}^2})\\\label{90b}&&\times\frac{\dot
{\textsf{Z}}}{{\textsf{Z}}}+\frac{\mathfrak{s}^2}{\textsf{Z}^4}+\frac{\mathfrak{s}^2\mathfrak{s}'}{\textsf{Z}^4\textsf{Y}}
+\frac{4\mathfrak{s}\dot{\mathfrak{s}}}{\textsf{Z}^4}-\frac{\mathfrak{s}^2\dot{\textsf{Z}}}{2\textsf{Z}^5}.
\end{eqnarray}

The stability of ${\mathbb{Y}}_{\mathcal{T}F}=0$ is governed by the
energy density and pressure. It can be noticed that the above
equation shows how the involvement of charge, energy density
inhomogeneity, pressure anisotropy and modified terms cause the
system to diverge from its steady state. It is also seen that the
contribution of electromagnetic field disturbs the stable character.
Moreover, the derivative of Eq.\eqref{87a} along with the
constraints ${\mathbb{Y}}_{\mathcal{T}F}=\sigma=0$ produces
\begin{align}\nonumber
&-\ddot
{\mathbb{Y}}_{\mathcal{T}F}-4\pi\ddot\Pi^{\textsf{(cor)}}+\left[4\pi\left(\frac{\mathcal{T}^{00\textsf{(cor)}}}{{\textsf{Y}}^2}\right)'\right]^.
-\ddot
{\textsf{L}}^{\textsf{(cor)}}-4\pi\left(\mathcal{T}^{\textsf{(cor)}}_{00}\right)^{..}+\dot
{\textsf{L}}^{\textsf{(cor)}}-4\pi \dot
{\texttt{B}_{1}}\\\nonumber&-12\pi\big\{\big(
\frac{\mathfrak{s}^2}{8\pi\textsf{Z}^4}+\mathcal{T}^{00\textsf{(cor)}}\big)\frac{\dot
{\textsf{Z}}}{{\textsf{Z}}}\big\}^.+\frac{3\dot
{\textsf{Z}}}{{\textsf{Z}}}[4\pi\left(\mathcal{T}^{\textsf{(cor)}}_{00}\right)^{.}-4\pi\left(\frac{\mathcal{T}^{00\textsf{(cor)}}}{{\textsf{Y}}^2}\right)'
+4\pi\dot\Pi^{\textsf{(cor)}}\\\nonumber&+4\pi
\texttt{B}_{1}+\frac{3\mathfrak{s}\dot{\mathfrak{s}}}{\textsf{Z}^4}-4\pi\left(\frac{{\textsf{Y}}'\mathcal{T}^{00\textsf{(cor)}}}
{{\textsf{Y}}^2}\right)-3{\textsf{L}}^{\textsf{(cor)}}\frac{\dot
{\textsf{Z}}}{{\textsf{Z}}}+12\pi\left(\frac{\mathcal{T}^{00\textsf{(cor)}}\dot
{\textsf{Z}}}{{\textsf{Z}}}\right)+8\pi\dot\Pi
-\frac{5\mathfrak{s}^2\dot{\textsf{Z}}}{2\textsf{Z}^5}\\\nonumber&-\left(12\pi
\mathcal{T}^{\textsf{(cor)}}_{00}+\frac{12\pi
\mathcal{T}^{\textsf{(cor)}}_{11}}{{\textsf{Y}}^2}\right)\frac{\dot
{\textsf{Z}}}{{\textsf{Z}}}+\frac{\mathfrak{s}^2}{\textsf{Z}^4}+\frac{\mathfrak{s}^2\mathfrak{s}'}{\textsf{Z}^4\textsf{Y}}
]+3\left(\frac{{\textsf{L}}^{\textsf{(cor)}}\dot
{\textsf{Z}}}{{\textsf{Z}}}\right)^.-\frac{16\pi\Pi^{\textsf{(tot)}}\dot
{\textsf{Z}}}{{\textsf{Z}}}-8\pi\ddot\Pi\\\nonumber&+\left[(12\pi
\mathcal{T}^{\textsf{(cor)}}_{00}+\frac{12\pi
\mathcal{T}^{\textsf{(cor)}}_{11}}{{\textsf{Y}}^2})\frac{\dot
{\textsf{Z}}}{{\textsf{Z}}}\right]^.+4\pi\left(\frac{{\textsf{Y}}'\mathcal{T}^{00\textsf{(cor)}}}{{\textsf{Y}}^2}\right)^.
+\big(\frac{\mathfrak{s}^2}{\textsf{Z}^4}+\frac{\mathfrak{s}^2\mathfrak{s}'}{\textsf{Z}^4\textsf{Y}}+\frac{3\mathfrak{s}\dot{\mathfrak{s}}}{\textsf{Z}^4}
-\frac{5\mathfrak{s}^2\dot{\textsf{Z}}}{2\textsf{Z}^5}\big)^.=0.\\\label{89a}
\end{align}
It is worth noting that along with other physical factors, heat
dissipation is assumed to be an extra component in the dissipative
case to disturb the zero complexity condition.

\section{Final Remarks}

The primary goal of this paper is to assess the complexity of
charged anisotropic dynamical cylinder in the context of non-minimal
coupled theory. The internal regime possesses anisotropic pressure,
charge, inhomogeneous energy density and heat flux. We have split
the Riemann tensor orthogonally by employing Herrera's method which
led to four scalar functions. Each of these scalars reveals some
specific physical aspects of the system. We have selected
${\mathbb{Y}}_{\mathcal{T}F}$ out of four scalars for inducing
complexity within the system because of the following reasons.
\begin{enumerate}
\item It has been found that this scalar is assumed to be the complexity producing
factor for the static structure in the context of GR as well as
$f(G,\mathcal{T})$ theory. So, by utilizing Eq.\eqref{43a}, we can
regain this scalar function for the static symmetric geometry from
the radiating anisotropic structure.
\item This factor encompasses all of the physical quantities including heat flow,
anisotropic pressure, charge, inhomogeneous energy density and extra
curvature terms, which make the system more complicated.
\end{enumerate}
For the present setup, homologous and homogeneous are the two
distinct evolutionary patterns that have been examined. We have
evaluated the findings for dissipative and non-dissipative models by
employing the vanishing complexity criterion and identifying the
homologous pattern as the most fundamental evolutionary process.
During the evolution phase, the parameters responsible for the
deviation of the system from the null complexity zone were also
discussed.

The extra curvature factors of $f(G,\mathcal{T})$ theory and matter
variables manifest significant influence on the substantial
characteristics of charged cylindrical source. This shows that the
involvement of modified terms puts forward a more complicated
dynamical cylinder. The geodesic nature, i.e., $\textsf{X}=1$ of the
homologous matter source during the evolution represents it to be
the simplest mode and includes extra terms of this theory. To
achieve a complex free configuration in GR, we have to impose the
constraints $\Pi={\mathcal{U}}={\varphi}=0$. However, an additional
condition,
${\textsf{L}}^{\textsf{(cor)}}-\frac{\mathfrak{s}^2}{\textsf{Z}^4}+\frac{\Pi}{2}f_{\mathcal{T}}-4\pi\Pi^{\textsf{(cor)}}=0$,
along with the above-mentioned factors must be prescribed to obtain
${\mathbb{Y}}_{\mathcal{T}F}=0$ (complex free system) for this
theory. The presence of shear even in the non-dissipative scenario
due to the involvement of modified terms led to the development of a
more complicated charged dynamical cylinder in contrast to GR. In
order to acquire the feasible solutions corresponding to the
dissipative as well as non-dissipative cases, the vanishing
complexity and homologous constraints have been utilized. It is
worth mentioning here that we have found the parameters that disturb
the stability of the zero complexity condition. Moreover, we have
observed that the system becomes more complicated in the presence of
electromagnetic field and additional curvature components of this
theory. Finally, the acquired results of this theory comply with GR
through the condition $f(G,\mathcal{T})=0$.

\section*{Appendix A}
\renewcommand{\theequation}{A\arabic{equation}}
\setcounter{equation}{0} The correction terms of $f(G,\mathcal{T})$
theory are expressed as
\begin{eqnarray}\nonumber
\mathcal{T}^{\textsf{(cor)}}_{00}&=&\frac{1}{8\pi}\left[\left({\mathcal{U}}+{{\mathcal{P}}}\right){\textsf{X}}^2f_{\mathcal{T}}-\frac{{\textsf{X}}^2}{2}f
+\left(\frac{8\dot{{\textsf{Y}}}\dot{{\textsf{Z}}}\ddot{{\textsf{Z}}}}{{\textsf{X}}^2{\textsf{Y}}
{\textsf{Z}}^2}-\frac{16{\textsf{X}}'\dot {\textsf{Z}}\dot
{\textsf{Z}}'}{{\textsf{X}}{\textsf{Y}}^2{\textsf{Z}}^2}+\frac{8{\textsf{X}}{\textsf{Z}}'{\textsf{X}}'
{\textsf{Z}}''}{{\textsf{Y}}^4{\textsf{Z}}^2}\right.\right.\\\nonumber
&+&\left.\left. \frac{4\ddot
{\textsf{Y}}}{{\textsf{Y}}{\textsf{Z}}^2}+\frac{8{\textsf{X}}'\dot
{\textsf{Y}}\dot
{\textsf{Z}}{\textsf{Z}}'}{{\textsf{X}}{\textsf{Y}}^3{\textsf{Z}}^2}-\frac{8\dot
{\textsf{X}}{\textsf{Y}}'{\textsf{Z}}'\dot
{\textsf{Z}}}{{\textsf{X}}{\textsf{Y}}^3{\textsf{Z}}^2}+\frac{8\dot
{\textsf{X}}\dot
{\textsf{Z}}{\textsf{Z}}''}{{\textsf{X}}{\textsf{Y}}^2{\textsf{Z}}^2}+\frac{4\dot
{\textsf{Z}}^2{\textsf{X}}'}{{\textsf{X}}{\textsf{Z}}^2}-\frac{12\dot
{\textsf{X}}\dot {\textsf{Y}}\dot
{\textsf{Z}}^2}{{\textsf{X}}^3{\textsf{Y}}{\textsf{Z}}^2}\right.\right.\\\nonumber
&\times&\left.\left.\frac{{\textsf{Y}}'}{{\textsf{Y}}^3}+\frac{4{\dot
{\textsf{X}}\textsf{Z}}'^2\dot
{\textsf{Y}}}{{\textsf{X}}{\textsf{Y}}^3{\textsf{Z}}^2}-\frac{12{\textsf{X}}{\textsf{X}}'{\textsf{Y}}'{\textsf{Z}}'^2}
{{\textsf{Y}}^5{\textsf{Z}}^2}+\frac{4{\textsf{X}}{\textsf{Z}}'^2{\textsf{X}}''}{{\textsf{Y}}^4{\textsf{Z}}^2}+\frac{4\dot
{\textsf{Z}}^2\ddot {\textsf{Y}}
}{{\textsf{X}}^2{\textsf{Y}}{\textsf{Z}}^2}+\frac{8\dot
{\textsf{Z}}'^2}{{\textsf{Y}}^2{\textsf{Z}}^2}\right.\right.\\\nonumber
&+&\left.\left.\frac{8{\textsf{Y}}'\ddot
{\textsf{Z}}{\textsf{Z}}'}{{\textsf{Y}}^3{\textsf{Z}}^2}-\frac{16\dot
{\textsf{Y}}{\textsf{Z}}'\dot
{\textsf{Z}}'}{{\textsf{Z}}^2{\textsf{Y}}^3}-\frac{4\dot
{\textsf{X}}\dot
{\textsf{Y}}}{{\textsf{X}}{\textsf{Z}}^2{\textsf{Y}}}-\frac{4\dot
{\textsf{Z}}^2{\textsf{X}}''}{{\textsf{Y}}^2{\textsf{X}}{\textsf{Z}}^2}+\frac{8\dot
{\textsf{Z}}^2{\textsf{X}}'^2}{{\textsf{Y}}^2{\textsf{X}}^2{\textsf{Z}}^2}+\frac{8{\textsf{Z}}'^2\dot
{\textsf{Y}}^2}{{\textsf{Z}}^2{\textsf{Y}}^4}\right.\right.\\\nonumber
&-&\left.\left.\frac{8\ddot
{\textsf{Z}}{\textsf{Z}}''}{{\textsf{Z}}^2{\textsf{Y}}^2}-\frac{4{\textsf{X}}{\textsf{X}}''}{{\textsf{Z}}^2{\textsf{Y}}^2}
+\frac{{\textsf{Y}}'{\textsf{X}}{\textsf{X}}'}{{\textsf{Y}}^3{\textsf{Z}}^2}
-\frac{4{\textsf{Z}}'^2\ddot
{\textsf{Y}}}{{\textsf{Y}}^3{\textsf{Z}}^2}\right)f_{G}+\left(\frac{8\dot
{\textsf{Z}}{\textsf{Z}}''}{{\textsf{Y}}^2{\textsf{Z}}^2}-\frac{8{\textsf{Y}}'{\textsf{Z}}'\dot
{\textsf{Z}}}{{\textsf{Y}}^3{\textsf{Z}}^2}\right.\right.\\\nonumber
&-&\left.\left.\frac{4\dot
{\textsf{Y}}}{{\textsf{Y}}{\textsf{Z}}^2}-\frac{12\dot
{\textsf{Y}}\dot
{\textsf{Z}}^2}{{\textsf{X}}^2{\textsf{Y}}{\textsf{Z}}^2}+\frac{4\dot
{\textsf{Y}}{\textsf{Z}}'^2}{{\textsf{Y}}^3{\textsf{Z}}^2}\right)\dot
f_{G}+\left(\frac{8\dot {\textsf{Y}} \dot
{\textsf{Z}}{\textsf{Z}}'}{{\textsf{Y}}^3{\textsf{Z}}^2}-\frac{4{\textsf{Y}}'\dot
{\textsf{Z}}^2}{{\textsf{Y}}^3{\textsf{Z}}^2}-\frac{4{\textsf{X}}^2{\textsf{Y}}'}{{\textsf{Y}}^3{\textsf{Z}}^2}\right.\right.
\\\nonumber&+&\left.\left.\frac{12{\textsf{X}}^2{\textsf{Y}}'{\textsf{Z}}'^2}{{\textsf{Y}}^5{\textsf{Z}}^2}
-\frac{8{\textsf{X}}^2{\textsf{Z}}'{\textsf{Z}}''}{{\textsf{Y}}^4{\textsf{Z}}^2}\right)f'_{G}
+\left(\frac{4\dot
{\textsf{Z}}^2}{{\textsf{Y}}^2{\textsf{Z}}^2}+\frac{4{\textsf{X}}^2}{{\textsf{Y}}^2{\textsf{Z}}^2}
-\frac{4{\textsf{X}}^2{\textsf{Z}}'^2}{{\textsf{Y}}^4{\textsf{Z}}^2}\right)f''_{G}\right],\\\label{100}
\\\nonumber
\mathcal{T}^{\textsf{(cor)}}_{01}&=&\frac{1}{8\pi}\left[-{\varphi}{\textsf{X}}{\textsf{Y}}f_{\mathcal{T}}
+\left(\frac{10{\textsf{X}}'{\textsf{Y}}'{\textsf{Z}}'\dot
{\textsf{Y}}}{{\textsf{X}}{\textsf{Y}}^4{\textsf{Z}}}-\frac{10\dot
{\textsf{X}}\dot {\textsf{Y}}\dot
{\textsf{Z}}{\textsf{X}}'}{{\textsf{X}}^4{\textsf{Y}}{\textsf{Z}}}-\frac{8\dot
{\textsf{X}}\dot {\textsf{Y}}\dot
{\textsf{Z}}'}{{\textsf{X}}^3{\textsf{Y}}{\textsf{Z}}^2}-\frac{8{\textsf{X}}'
}{{\textsf{X}}{\textsf{Y}}^3}\right.\right.\\\nonumber
&\times&\left.\left.\frac{\dot
{\textsf{Y}}{\textsf{Z}}'^2}{{\textsf{Z}}^2}-\frac{10\dot
{\textsf{X}}\dot
{\textsf{Y}}^2{\textsf{Z}}'}{{\textsf{X}}^3{\textsf{Y}}^2{\textsf{Z}}}-\frac{8{\textsf{X}}'^2{\textsf{Z}}'\dot
{\textsf{Z}}}{{\textsf{X}}^2{\textsf{Y}}^2{\textsf{Z}}^2}+\frac{10{\textsf{X}}'^2{\textsf{Y}}'\dot
{\textsf{Z}}}{{\textsf{X}}^2{\textsf{Y}}^3{\textsf{Z}}}+\frac{\dot
{\textsf{Y}}\ddot
{\textsf{Z}}{\textsf{Z}}'}{{\textsf{X}}^2{\textsf{Y}}{\textsf{Z}}^2}+\frac{10\dot
{\textsf{Y}} \ddot
{\textsf{Y}}{\textsf{Z}}'}{{\textsf{X}}^2{\textsf{Y}}^2{\textsf{Z}}}\right.\right.\\\nonumber
&+&\left.\left.\frac{10{\textsf{X}}'\ddot {\textsf{Y}}\dot
{\textsf{Z}}}{{\textsf{X}}^3{\textsf{Y}}{\textsf{Z}}}-\frac{10{\textsf{X}}''{\textsf{X}}'\dot
{\textsf{Z}}}{{\textsf{X}}^2{\textsf{Y}}^2{\textsf{Z}}}-\frac{10{\textsf{X}}''{\textsf{Z}}'\dot
{\textsf{Y}}}{{\textsf{X}}{\textsf{Y}}^3{\textsf{Z}}}+\frac{8{\textsf{X}}'{\textsf{Z}}'}{{\textsf{X}}{\textsf{Y}}^2}+\frac{\dot
{\textsf{X}}\dot {\textsf{Y}}\dot
{\textsf{Z}}'}{{\textsf{X}}^3{\textsf{Y}}{\textsf{Z}}}-\frac{10{\textsf{X}}'\dot
{\textsf{Z}}'{\textsf{Y}}'\dot
{\textsf{Z}}'}{{\textsf{X}}{\textsf{Y}}^3{\textsf{Z}}{\textsf{Z}}^2}\right.\right.\\\nonumber
&+&\left.\left.\frac{8{\textsf{X}}'\dot {\textsf{Z}}\ddot
{\textsf{Z}}}{{\textsf{X}}^3{\textsf{Z}}^2}-\frac{8{\textsf{X}}'\dot
{\textsf{X}}\dot
{\textsf{Z}}^2}{{\textsf{X}}^4{\textsf{Z}}^2}+\frac{8\dot
{\textsf{X}}\dot {\textsf{Z}}\dot
{\textsf{Z}}'}{{\textsf{X}}^3{\textsf{Z}}^2}-\frac{10\ddot
{\textsf{Y}}\dot
{\textsf{Z}}'}{{\textsf{X}}^2{\textsf{Y}}{\textsf{Z}}}+\frac{10{\textsf{X}}''\dot
{\textsf{Z}}'}{{\textsf{X}}{\textsf{Y}}^2{\textsf{Z}}}-\frac{8\ddot
{\textsf{Z}}\dot
{\textsf{Z}}'}{{\textsf{X}}^2{\textsf{Z}}^2}\right)f_{G}\right.\\\nonumber&+&\left.\left(-\frac{8\dot
{\textsf{X}}\dot
{\textsf{Y}}{\textsf{X}}'}{{\textsf{X}}^4{\textsf{Y}}}-\frac{4{\textsf{X}}'}{{\textsf{X}}{\textsf{Z}}^2}+\frac{8\dot
{\textsf{Z}}\dot
{\textsf{Z}}'}{{\textsf{X}}^2{\textsf{Z}}^2}+\frac{4{\textsf{X}}'{\textsf{Z}}'^2}{{\textsf{X}}{\textsf{Y}}^2{\textsf{Z}}^2}
+\frac{8{\textsf{X}}'^2{\textsf{Y}}'}{{\textsf{X}}^2{\textsf{Y}}^3}-\frac{8{\textsf{X}}'{\textsf{X}}''}{{\textsf{X}}^2{\textsf{Y}}^2}+\frac{8{\textsf{X}}'\ddot
{\textsf{Y}}}{{\textsf{X}}^3{\textsf{Y}}}\right.\right.\\\nonumber&-&\left.\left.\frac{12{\textsf{X}}'\dot
{\textsf{Z}}^2}{{\textsf{X}}^3{\textsf{Z}}^2}\right)\dot
f_{G}+\left(\frac{8{\textsf{X}}'{\textsf{Z}}'\dot
{\textsf{Z}}}{{\textsf{X}}{\textsf{Y}}^2{\textsf{Z}}^2}+\frac{8{\textsf{X}}'{\textsf{Y}}'\dot
{\textsf{Y}}}{{\textsf{X}}{\textsf{Y}}^4}-\frac{4\dot
{\textsf{Y}}\dot
{\textsf{Z}}^2}{{\textsf{X}}^2{\textsf{Y}}{\textsf{Z}}^2}+\frac{12\dot
{\textsf{Y}}{\textsf{Z}}'^2}{{\textsf{Y}}^3{\textsf{Z}}^2}+\frac{8\ddot
{\textsf{Y}}\dot
{\textsf{Y}}}{{\textsf{X}}^2{\textsf{Y}}^2}\right.\right.\\\nonumber&-&\left.\left.\frac{8\dot
{\textsf{X}}\dot
{\textsf{Y}}^2}{{\textsf{X}}^3{\textsf{Y}}^2}-\frac{8{\textsf{X}}''\dot
{\textsf{Y}}}{{\textsf{X}}{\textsf{Y}}^3}-\frac{8{\textsf{Z}}'\dot
{\textsf{Z}}'}{{\textsf{Y}}^2{\textsf{Z}}^2}-\frac{4\dot
{\textsf{Y}}}{{\textsf{Y}}{\textsf{Z}}^2}\right)f'_{G}+\left(\frac{4\dot
{\textsf{Z}}^2}{{\textsf{X}}^2{\textsf{Z}}^2}-\frac{8{\textsf{X}}'{\textsf{Y}}'}{{\textsf{X}}{\textsf{Y}}^3}+\frac{8\dot
{\textsf{X}}\dot
{\textsf{Y}}}{{\textsf{X}}^3{\textsf{Y}}}\right.\right.
\\\label{100a}&+&\left.\left.\frac{4}{{\textsf{Z}}^2}-\frac{4{\textsf{Z}}'^2}{{\textsf{Y}}^2{\textsf{Z}}^2}
+\frac{8{\textsf{X}}''}{{\textsf{X}}{\textsf{Y}}^2}-\frac{8\ddot
{\textsf{Y}}}{{\textsf{X}}^2{\textsf{Y}}}\right)\dot
f'_{G}\right],\\\nonumber
\mathcal{T}^{\textsf{(cor)}}_{11}&=&\frac{1}{8\pi}\left[\frac{2}{3}\Pi
{\textsf{Y}}^2
f_{\mathcal{T}}+\frac{{\textsf{Y}}^2}{2}f+\left(\frac{32\dot
{\textsf{Y}}{\textsf{Z}}' \dot
{\textsf{Z}}'^2}{{\textsf{X}}^2{\textsf{Y}}{\textsf{Z}}^2}-\frac{8{\textsf{X}}'{\textsf{Z}}'{\textsf{Z}}''}{{\textsf{X}}{\textsf{Y}}^2{\textsf{Z}}^2}
-\frac{8{\textsf{Y}}\dot {\textsf{Y}}\dot {\textsf{Z}}\ddot
{\textsf{Z}}}{{\textsf{X}}^2{\textsf{Z}}^4}\right.\right.\\\nonumber
&-&\left.\left.\frac{8{\textsf{Y}}'\ddot
{\textsf{Z}}{\textsf{Z}}'}{{\textsf{Y}}{\textsf{Z}}^2{\textsf{X}}^2}+\frac{\ddot
{\textsf{Y}}{\textsf{Z}}'^2}{{\textsf{X}}^2{\textsf{Y}}{\textsf{Z}}^2}-\frac{4{\textsf{X}}''{\textsf{Z}}'^2}{{\textsf{X}}{\textsf{Y}}^2{\textsf{Z}}^2}
-\frac{16\dot
{\textsf{Z}}'^2}{{\textsf{X}}^2{\textsf{Z}}^2}-\frac{8\dot
{\textsf{X}}\dot
{\textsf{Z}}{\textsf{Z}}''}{{\textsf{X}}^3{\textsf{Z}}^2}+\frac{32{\textsf{X}}'\dot
{\textsf{Z}}\dot
{\textsf{Z}}'}{{\textsf{X}}^3{\textsf{Z}}^2}\right.\right.\\\nonumber
&-&\left.\left.\frac{4{\textsf{Y}}\ddot {\textsf{Y}}\dot
{\textsf{Z}}^2}{{\textsf{X}}^4{\textsf{Z}}^2}-\frac{\dot
{\textsf{Y}}^2{\textsf{Z}}'^2}{{\textsf{X}}^2{\textsf{Y}}^2{\textsf{Z}}^2}+\frac{8\ddot
{\textsf{Z}}{\textsf{Z}}''}{{\textsf{X}}^2{\textsf{Z}}^2}+\frac{4{\textsf{Y}}\dot
{\textsf{Y}}\dot
{\textsf{X}}}{{\textsf{X}}^3{\textsf{Z}}^2}-\frac{4{\textsf{X}}'{\textsf{Y}}'}{{\textsf{X}}{\textsf{Y}}{\textsf{Z}}^2}+\frac{4{\textsf{X}}''\dot
{\textsf{Z}}^2}{{\textsf{X}}^3{\textsf{Z}}^2}\right.\right.\\\nonumber
&-&\left.\left.\frac{24{\textsf{X}}'\dot {\textsf{Y}}\dot
{\textsf{Z}}{\textsf{Z}}'}{{\textsf{X}}^3{\textsf{Y}}{\textsf{Z}}^2}-\frac{4{\textsf{Y}}\ddot
{\textsf{Y}}}{{\textsf{X}}^2{\textsf{Z}}^2}+\frac{8\dot
{\textsf{X}}\dot
{\textsf{Z}}{\textsf{Y}}'{\textsf{Z}}'}{{\textsf{X}}^3{\textsf{Y}}{\textsf{Z}}^2}-\frac{16{\textsf{X}}'^2\dot
{\textsf{Z}}^2}{{\textsf{X}}^4{\textsf{Z}}^2}+\frac{4{\textsf{X}}''}{{\textsf{X}}{\textsf{Z}}^2}
+\frac{12{\textsf{X}}'{\textsf{Z}}'^2{\textsf{Y}}'}{{\textsf{X}}{\textsf{Y}}^3
{\textsf{Z}}^2}\right.\right.\\\nonumber
&-&\left.\left.\frac{4{\textsf{X}}'{\textsf{Y}}'\dot
{\textsf{Z}}^2}{{\textsf{X}}^3{\textsf{Y}}{\textsf{Z}}^2}-\frac{4\dot
{\textsf{X}}\dot
{\textsf{Y}}{\textsf{Z}}'^2}{{\textsf{X}}^3{\textsf{Y}}{\textsf{Z}}^2}\right)f_{G}+\left(\frac{8{\textsf{Y}}^2\dot
{\textsf{Z}}\ddot
{\textsf{Z}}}{{\textsf{X}}^4{\textsf{Z}}^2}-\frac{12{\textsf{Y}}^2\dot
{\textsf{X}}\dot
{\textsf{Z}}^2}{{\textsf{X}}^5{\textsf{Z}}^2}+\frac{4\dot
{\textsf{X}}{\textsf{Z}}'^2}{{\textsf{X}}^3{\textsf{Z}}^2}\right.\right.\\\nonumber
&-&\left.\left.\frac{8{\textsf{X}}'\dot
{\textsf{Z}}{\textsf{Z}}'}{{\textsf{X}}^3{\textsf{Z}}^2}-\frac{4{\textsf{Y}}^2\dot
{\textsf{X}}}{{\textsf{X}}^3{\textsf{Z}}^2}\right)\dot
f_{G}+\left(\frac{4{\textsf{Y}}^2\dot
{\textsf{Z}}^2}{{\textsf{X}}^4{\textsf{Z}}^2}\frac{4{\textsf{Y}}^2}{{\textsf{X}}^2{\textsf{Z}}^2}
-\frac{4{\textsf{Z}}'^2}{{\textsf{X}}^2{\textsf{Z}}^2}\right)\ddot
f_{G}+\left(\frac{8\dot {\textsf{Z}}\dot
{\textsf{X}}{\textsf{Z}}'}{{\textsf{X}}^3{\textsf{Z}}^2}\right.\right.\\\label{100b}
&-&\left.\left.\frac{4\dot
{\textsf{X}}'{\textsf{Z}}^2}{{\textsf{X}}^3{\textsf{Z}}^2}-\frac{4{\textsf{X}}'}{{\textsf{X}}{\textsf{Z}}^2}
+\frac{12{\textsf{Z}}'^2{\textsf{X}}'}{{\textsf{X}}{\textsf{Y}}^2{\textsf{Z}}^2}-\frac{\ddot
{\textsf{Z}}{\textsf{Z}}'}{{\textsf{X}}^2{\textsf{Z}}^2}\right)f'_{G}\right],\\\nonumber
\mathcal{T}^{\textsf{(cor)}}_{22}&=&\frac{1}{8\pi}\left[\frac{
-{\textsf{Z}}^2 }{3}\Pi
f_{\mathcal{T}}+\frac{{\textsf{Z}}^2}{2}f+\left(\frac{-4\dot
{\textsf{Z}}^2}{{\textsf{X}}^2{\textsf{Y}}^2}+\frac{4{\textsf{X}}''}{{\textsf{X}}{\textsf{Y}}^2}+\frac{8\dot
{\textsf{X}}\dot
{\textsf{Z}}{\textsf{Y}}'{\textsf{Z}}'}{{\textsf{X}}^3{\textsf{Y}}^3}-\frac{4\ddot
{\textsf{Y}}}{{\textsf{X}}^2{\textsf{Y}}}\right.\right.\\\nonumber&-&\left.\left.\frac{4\dot
{\textsf{X}}\dot
{\textsf{Y}}{\textsf{Z}}'^2}{{\textsf{Y}}^3{\textsf{X}}^3}-\frac{4{\textsf{X}}'{\textsf{Y}}'\dot
{\textsf{Z}}^2}{{\textsf{Y}}^3{\textsf{X}}^3}+\frac{12\dot
{\textsf{Y}}\dot {\textsf{X}}\dot
{\textsf{Z}}^2}{{\textsf{Y}}{\textsf{X}}^5}+\frac{12{{\textsf{Y}}'\textsf{X}}'{\textsf{Z}}'^2}{{\textsf{X}}{\textsf{Y}}^5}
-\frac{8{\textsf{X}}'\textsf{Z}
'{\textsf{Z}}''}{{\textsf{Y}}^4{\textsf{X}}}-\frac{4{\textsf{X}}'{\textsf{Y}}'}{{\textsf{X}}{\textsf{Y}}^3}
\right.\right.\\\nonumber&-&\left.\left.\frac{8\dot {\textsf{X}}\dot
{\textsf{Z}}{\textsf{Z}}''}{{\textsf{X}}^3{\textsf{Y}}^2}
-\frac{8\dot {\textsf{Y}}\dot {\textsf{Z}} \ddot
{\textsf{Z}}}{{\textsf{X}}^4{\textsf{Y}}}-\frac{8{\textsf{Z}}'{\textsf{Y}}'\ddot
{\textsf{Z}}}{{\textsf{Y}}^3{\textsf{X}}^2}+\frac{8{\textsf{X}}'\dot
{\textsf{Z}}\dot
{\textsf{Z}}'}{{\textsf{X}}^3{\textsf{Y}}^2}+\frac{4\dot
{\textsf{X}}\dot
{\textsf{Y}}}{{\textsf{X}}^3{\textsf{Y}}}-\frac{4{{\textsf{Z}}'^2\textsf{X}}''}{{\textsf{X}}{\textsf{Y}}^4}
\right.\right.\\\nonumber&-&\left.\left.\frac{4{\textsf{X}}'^2 \dot
{\textsf{Z}}^2}{{\textsf{X}}^4{\textsf{Y}}^2}-\frac{4\ddot
{\textsf{Y}} \dot {\textsf{Z}}^2}{{\textsf{X}}^4
{\textsf{Y}}}+\frac{4\ddot
{\textsf{Y}}{\textsf{Z}}'^2}{{\textsf{X}}^2{\textsf{Y}}^3}+\frac{4{\textsf{X}}''\dot
{\textsf{Z}}^2}{{\textsf{X}}^3{\textsf{Y}}^2}+\frac{8\ddot
{\textsf{Z}}{\textsf{Z}}''}{{\textsf{X}}^2{\textsf{Y}}^2}-\frac{4\dot
{\textsf{Y}}^2{\textsf{Z}}'^2}{{\textsf{X}}^2{\textsf{Y}}^4}\right)f_{G}\right.\\\nonumber&+&\left.\left(\frac{4\ddot
{\textsf{Y}}{\textsf{Z}}\dot
{\textsf{Z}}}{{\textsf{X}}^4{\textsf{Y}}}-\frac{12\dot
{\textsf{X}}\dot {\textsf{Y}} \dot
{\textsf{Z}}{\textsf{Z}}}{{\textsf{X}}^5{\textsf{Y}}}-\frac{4\dot
{\textsf{X}}{\textsf{Y}}'{\textsf{Z}}
{\textsf{Z}}'}{{\textsf{X}}^3{\textsf{Y}}^3}+\frac{8{\textsf{X}}'{\textsf{Z}}'\dot
{\textsf{Y}}}{{\textsf{X}}^3{\textsf{Y}}^3}+\frac{4{\textsf{X}}'{\textsf{Y}}'\dot
{\textsf{Z}}{\textsf{Z}}}{{\textsf{X}}^3{\textsf{Y}}^3}+\frac{4\dot
{\textsf{Y}}\ddot {\textsf{Z}}{\textsf{Z}}
}{{\textsf{X}}^4{\textsf{Y}}}\right.\right.\\\nonumber&-&\left.\left.\frac{4{\textsf{X}}''\dot
{\textsf{Z}}{\textsf{Z}}}{{\textsf{Y}}^2{\textsf{X}}^3}-\frac{12{\textsf{X}}'{\textsf{Z}}\dot
{\textsf{Z}}'}{{\textsf{X}}^3{\textsf{Y}}^2}+\frac{4\dot
{\textsf{X}}{\textsf{Z}}{\textsf{Z}}''}{{\textsf{X}}^3{\textsf{Y}}^2}+\frac{12{\textsf{Z}}\dot
{\textsf{Z}}{\textsf{X}}'^2}{{\textsf{X}}^4{\textsf{Y}}^2}\right)\dot
f_{G}+\left(\frac{8{\textsf{X}}'\dot {\textsf{Z}}\dot
{\textsf{Y}}{\textsf{Z}}}{{\textsf{X}}^3{\textsf{Y}}^3}\right.\right.\\\nonumber&+&\left.\left.\frac{4{\textsf{Y}}'{\textsf{Z}}\ddot
{\textsf{Z}}}{{\textsf{Y}}^3{\textsf{X}}^2}+\frac{4{\textsf{Z}}\dot
{\textsf{X}}{\textsf{Z}}'\dot
{\textsf{Y}}}{{\textsf{X}}^3{\textsf{Y}}^3}-\frac{12{{\textsf{Z}}'{\textsf{Z}}\textsf{X}}'{\textsf{Y}}'}{{\textsf{X}}
{\textsf{Y}}^5}-\frac{4{\dot
{\textsf{Z}}{\textsf{Z}}\textsf{Y}}'\dot
{\textsf{X}}}{{\textsf{Y}}^3{\textsf{X}}^3}+\frac{4{{\textsf{Z}}{\textsf{Z}}'\textsf{X}}''}{{\textsf{X}}{\textsf{Y}}^4}
\right.\right.\\\nonumber&+&\left.\left.\frac{8\dot
{\textsf{Y}}{\textsf{Z}}'\dot
{\textsf{Z}}'}{{\textsf{X}}^2{\textsf{Y}}^3}-\frac{4\ddot
{\textsf{Y}}{\textsf{Z}}'{\textsf{Z}}}{{\textsf{Y}}^3{\textsf{X}}^2}-\frac{12\dot
{\textsf{Y}}\dot
{\textsf{Z}}'{\textsf{Z}}}{{\textsf{Y}}^3{\textsf{X}}^2}+\frac{4{{\textsf{Z}}\textsf{X}}'{\textsf{Z}}''}{{\textsf{Y}}^4{\textsf{X}}}
+\frac{12{\textsf{X}}'^2{\textsf{Z}}\dot
{\textsf{Z}}}{{\textsf{X}}^2{\textsf{Y}}^4}\right)f'_{G}\right.\\\nonumber&+&\left.\left(\frac{4\dot
{\textsf{X}}\dot
{\textsf{Z}}{\textsf{Z}}}{{\textsf{X}}^3{\textsf{Y}}^2}+\frac{4{\textsf{X}}'{\textsf{Z}}'{\textsf{Z}}}{{\textsf{X}}{\textsf{Y}}^4}-\frac{4\ddot
{\textsf{Z}}{\textsf{Z}}}{{\textsf{X}}^2{\textsf{Y}}^2}\right)f''_{G}+\left(\frac{12{\textsf{Z}}\dot
{\textsf{Z}}'}{{\textsf{X}}^2{\textsf{Y}}^2}-\frac{12{\textsf{Z}}\dot
{\textsf{Y}}{\textsf{Z}}'}{{\textsf{X}}^2{\textsf{Y}}^3}\right.\right.\\\label{100c}&-&\left.\left.\frac{12{\textsf{X}}'{\textsf{Z}}\dot
{\textsf{Z}}}{{\textsf{X}}^3{\textsf{Y}}^2}\right)\dot
f'_{G}+\left(\frac{4\dot {\textsf{Y}}\dot
{\textsf{Z}}{\textsf{Z}}}{{\textsf{X}}^4{\textsf{Y}}}+\frac{4{\textsf{Y}}'{\textsf{Z}}'{\textsf{Z}}}{{\textsf{X}}^2{\textsf{Y}}^3}
-\frac{4{\textsf{Z}}{\textsf{Z}}''}{{\textsf{X}}^2{\textsf{Y}}^2}\right)\ddot
f_{G}\right].
\end{eqnarray}

\section*{Appendix B}
\renewcommand{\theequation}{B\arabic{equation}}
\setcounter{equation}{0} The modified terms present in the structure
scalars are evaluated as
\begin{eqnarray*}
\textsf{W}^{\textsf{(cor)}}_{\eta\vartheta} &=&
2\left[R_{a\vartheta}R^{a}_{\eta}f_{G}-\frac{1}{2}RR_{\eta\vartheta}f_{G}+
R^{lm}R_{l\vartheta a\eta}f_{G}-\frac{1}{2}R_{\vartheta
lmn}R^{lmn}_{\eta}f_{G}+R_{\eta\vartheta}\Box
f_{G}\right.\\\nonumber&+&\left.\frac{1}{2}R
\nabla_{\eta}\nabla_{\vartheta}f_{G}-R^{a}_{\eta}\nabla_{\vartheta}\nabla_{a}f_{G}
-R^{a}_{\vartheta}\nabla_{\eta}\nabla_{a}f_{G}-R_{l\vartheta
a\eta}\nabla^{a}\nabla^{l}f_{G}\right]\\\nonumber
&+&4R^{lm}h_{\eta\vartheta}\nabla_{a}\nabla_{l}f_{G}-2Rh_{\eta\vartheta}\Box
f_{G}+2\left[-R_{a\delta}R^{a}_{\eta}f_{G}- R^{lm}R_{l\delta
a\eta}f_{G}\right.\\\nonumber&+&\left.\frac{1}{2}R_{\delta
lmn}R^{lmn}_{\eta}f_{G}+\frac{1}{2}RR_{\eta\vartheta}f_{G}-R_{\delta\eta}\Box
f_{G}+R_{l\delta
a\eta}\nabla^{a}\nabla^{a}f_{G}\right.\\\nonumber&-&\left.\frac{1}{2}R
\nabla_{\eta}\nabla_{\delta}f_{G}+R^{a}_{\eta}\nabla_{\delta}\nabla_{a}f_{G}
+R^{a}_{\delta}\nabla_{\eta}\nabla_{a}f_{G}\right]\mathcal{V}_{\vartheta}\mathcal{V}^{\delta}
+2\left[-R_{a\vartheta}R^{a\gamma}f_{G}\right.\\\nonumber&+&\left.\frac{1}{2}R_{\vartheta
lmn}R^{lmn\gamma}f_{G}- R^{lm}R^{\gamma}_{l\vartheta
a}f_{G}+\frac{1}{2}RR^{\gamma}_{\vartheta}f_{G}-R^{\gamma}_{\vartheta}\Box
f_{G}+R^{a\gamma}\nabla_{\vartheta}\nabla_{a}f_{G}\right.\\\nonumber&-&\left.\frac{1}{2}R
\nabla^{\gamma}\nabla_{\vartheta}f_{G}+R^{a}_{\vartheta}\nabla^{\gamma}\nabla_{a}f_{G}+R^{\gamma}_{a\vartheta
l}\nabla^{a}\nabla^{l}\right]\mathcal{V}_{\eta}\mathcal{V}_{\gamma}+
2\left[R_{a\delta}R^{a\gamma}f_{G}\right.\\\nonumber
&-&\left.\frac{1}{2}R_{\delta lmn}R^{lmn
\gamma}f_{G}+R^{lm}R^{\gamma}_{l\delta
a}f_{G}-\frac{1}{2}RR^{\gamma}_{\delta}f_{G}+R^{\gamma}_{\delta}\Box
f_{G}+\frac{1}{2}R
\nabla^{\gamma}\nabla_{\delta}f_{G}\right.\\\nonumber&-&\left.R^{\gamma}_{l\delta
a}\nabla^{a}\nabla^{l}f_{G}-R^{a\gamma}\nabla_{\delta}\nabla_{a}f_{G}-R^{a}_{\delta}\nabla^{\gamma}\nabla_{a}
f_{G}\right]
g_{\eta\vartheta}\mathcal{V}_{\gamma}\mathcal{V}^{\delta}
-\frac{1}{3}\left[-2R^{2}f_{G}\right.\\\nonumber&-&\left.2R^{\beta}_{lmn}R^{lmn}_{\beta}f_{G}
-2R\Box
f_{G}+4R^{a\alpha}R_{a\alpha}f_{G}+4R^{lm}R^{\alpha}_{a\alpha
l}f_{G}+16R^{lm}\nabla_{a}\nabla_{l}f_{G}\right.\\\nonumber&-&\left.4R^{a\beta}\nabla_{\beta}\nabla_{a}f_{G}
-4R^{a\alpha}\nabla_{\alpha}\nabla_{a}f_{G}-4R^{\alpha}_{l\alpha
a}\nabla^{a}\nabla^{l}f_{G}\right]h_{\eta\vartheta}
-\frac{1}{6}fh_{\eta\vartheta},
\end{eqnarray*}
\begin{eqnarray*}
\textsf{K}^{\textsf{(cor)}}_{\eta\vartheta}
&=&\left[\frac{1}{2}R_{a\epsilon}R^{a
p}f_{G}+\frac{1}{2}R^{lm}R^{p}_{a\epsilon
l}f_{G}-\frac{1}{4}RR^{p}_{\epsilon}f_{G}-\frac{1}{4}R_{\epsilon
lmn}R^{lmn p}f_{G}\right.\\\nonumber&+&\left.\frac{1}{2}
R^{p}_{\epsilon}\Box
f_{G}+\frac{1}{4}R\nabla^{p}\nabla_{\epsilon}f_{G}-\frac{1}{4}R^{lp}\nabla_{\epsilon}\nabla_{a}f_{G}-\frac{1}{2}R^{a}_{\epsilon}\nabla^{p}\nabla_{a}
f_{G}\right.\\\nonumber&-&\left.\frac{1}{2}R^{p}_{a\epsilon
l}\nabla^{a}\nabla^{l}f_{G}\right]\epsilon_{p
\delta\vartheta}\epsilon^{\epsilon\delta}_{\eta}+\left[-\frac{1}{2}R_{a\delta}R^{lp}f_{G}-\frac{1}{2}R^{lm}R^{p}_{a\delta
l}f_{G}\right.\\\nonumber&+&\left.\frac{1}{4}RR^{p}_{\delta}f_{G}+\frac{1}{4}R_{\delta
lmn}R^{lmnp}f_{G}-\frac{1}{2}R^{p}_{\delta}\Box
f_{G}-\frac{1}{4}R\nabla^{p}\nabla_{\delta}f_{G}\right.\\\nonumber&+&\left.\frac{1}{4}R^{lp}\nabla_{\delta}\nabla_{a}f_{G}
+\frac{1}{2}R^{a}_{\delta}\nabla^{p}\nabla_{a}f_{G}
+\frac{1}{2}R^{p}_{a\delta
l}\nabla^{a}\nabla^{l}f_{G}\right]\epsilon_{p\epsilon
\vartheta}\epsilon^{\epsilon\delta}_{\eta}\\\nonumber&+&\left[-\frac{1}{2}R_{a\epsilon}R^{a\gamma}f_{G}-\frac{1}{2}R^{lm}R^{\gamma}_{a\epsilon
l}f_{G}
+\frac{1}{4}RR^{\gamma}_{\epsilon}f_{G}+\frac{1}{4}R_{\epsilon
lmn}R^{lmn\gamma}f_{G}\right.\\\nonumber&-&\left.\frac{1}{2}R^{\gamma}_{\epsilon}\Box
f_{G}-\frac{1}{4}R\nabla^{\gamma}\nabla_{\epsilon}f_{G}+\frac{1}{4}R^{a\gamma}\nabla_{\epsilon}\nabla_{a}f_{G}
+\frac{1}{2}R^{a}_{\epsilon}\nabla^{\gamma}\nabla_{a}
f_{G}\right.\\\nonumber&+&\left.\frac{1}{2}R^{\gamma}_{a\epsilon
l}\nabla^{a}
\nabla^{l}f_{G}\right]\epsilon_{\delta\gamma\vartheta}\epsilon^{\epsilon\delta}_{\eta}+\left[\frac{1}{2}R_{a\delta}R^{a\gamma}f_{G}
+\frac{1}{2}R^{lm}R^{\gamma}_{a\delta
l}f_{G}\right.\\\nonumber&-&\left.\frac{1}{4}RR^{\gamma}_{\delta}f_{G}-\frac{1}{4}R_{\delta
lmn}R^{lmn\gamma}f_{G}+\frac{1}{2}R^{\gamma}_{\delta}\Box
f_{G}+\frac{1}{4}R\nabla^{\gamma}\nabla_{\delta}f_{G}\right.\\\nonumber&-&\left.\frac{1}{4}R^{a\gamma}\nabla_{\delta}\nabla_{a}f_{G}
-\frac{1}{2}R^{a}_{\delta}\nabla^{\gamma}\nabla_{a}f_{G}-\frac{1}{2}R^{\gamma}_{a\delta
l}\nabla^{a}\nabla^{l}f_{G}\right]\epsilon_{\epsilon\gamma\vartheta}\epsilon^{\epsilon\delta}_{\eta}\\\nonumber&-&4R^{lm}\nabla_{a}\nabla_{l}h_{\eta\vartheta}f_{G}
+2Rh_{\eta\vartheta}\Box f_{G}
+\frac{1}{3}\left[\left(\mathcal{U}+\mathcal{P}\right)f_{\mathcal{T}}+4R^{a\alpha}R_{a\alpha}f_{G}\right.\\\nonumber&+&\left.4R^{lm}R^{\alpha}_{l\alpha
a}f_{G}-2R^{2}f_{G}-2R^{\beta}_{lmn}R^{lmn}_{\beta}f_{G}-2R\Box
f_{G}\right.\\\nonumber&+&\left.16R^{lm}\nabla_{a}\nabla_{l}f_{G}-4R^{a\alpha}\nabla_{\alpha}\nabla_{a}f_{G}
-4R^{a\beta}\nabla_{\beta}\nabla_{a}f_{G}\right.\\\nonumber&-&\left.4R^{\alpha}_{l\alpha
a}\nabla^{a}\nabla^{l}f_{G}\right]h_{\eta\vartheta}+\frac{1}{6}fh_{\eta\vartheta},
\end{eqnarray*}
\begin{eqnarray}\nonumber
\textsf{M}^{\textsf{(cor)}}&=&
2\left[R_{\mu\vartheta}R^{\mu}_{\eta}f_{G}+
R^{\mu\nu}R_{\mu\vartheta\nu\eta}f_{G}-\frac{1}{2}RR_{\eta\vartheta}f_{G}-\frac{1}{2}R_{\vartheta\mu\nu
n}R^{\mu\nu n}_{\eta}f_{G}\right.\\\nonumber&+&\left.\frac{1}{2}R
\nabla_{\eta}\nabla_{\vartheta}f_{G}+R_{\eta\vartheta}\Box
f_{G}-R^{\mu}_{\eta}\nabla_{\vartheta}\nabla_{\mu}f_{G}-R^{\mu}_{\vartheta}\nabla_{\eta}\nabla_{\mu}f_{G}\right.\\\nonumber
&-&\left.R_{\mu\vartheta
\nu\eta}\nabla^{\mu}\nabla^{\nu}f_{G}\right]g^{\eta\vartheta}+2\left[-R_{\mu\delta}R^{\mu}_{\eta}f_{G}-
R^{\mu\nu}R_{\mu\delta
\nu\eta}f_{G}+\frac{1}{2}RR_{\eta\delta}f_{G}\right.\\\nonumber&+&\left.\frac{1}{2}R_{\delta
\mu\nu n}R^{\mu\nu n}_{\eta}f_{G}-R_{\delta\eta}\Box
f_{G}+R_{\mu\delta\nu\eta}\nabla^{\nu}\nabla^{\mu}f_{G}+R^{\mu}_{\eta}\nabla_{\mu}\nabla_{\delta}f_{G}
\right.\\\nonumber&+&\left.R^{\mu}_{\delta}\nabla_{\eta}\nabla_{\mu}
f_{G}-\frac{1}{2}R
\nabla_{\eta}\nabla_{\delta}f_{G}\right]{\mathcal{V}}_{\vartheta}{\mathcal{V}}^{\delta}g^{\eta\vartheta}+2\left[-R_{\mu\vartheta}R^{\mu\gamma}f_{G}\right.\\\nonumber&+&\left.
\frac{1}{2}R_{\vartheta\mu\nu n}R^{\mu\nu
n\gamma}f_{G}-+\frac{1}{2}RR^{\gamma}_{\vartheta}f_{G}
R^{\mu\nu}R^{\gamma}_{\mu\vartheta
\nu}f_{G}+R^{\mu\gamma}\nabla_{\vartheta}\nabla_{\mu}f_{G}\right.\\\nonumber&-&\left.\frac{1}{2}R
\nabla^{\gamma}\nabla_{\vartheta}f_{G}+R^{\gamma}_{\mu\vartheta
\nu}\nabla^{\nu}\nabla^{\mu}f_{G}-R^{\gamma}_{\vartheta}\Box
f_{G}+R^{\mu}_{\vartheta}\nabla_{\mu}\nabla^{\gamma}f_{G}\right]{\mathcal{V}}_{\eta}{\mathcal{V}}_{\gamma}g^{\eta\vartheta}\\\nonumber
&+&2\left[R^{\mu\gamma}R_{\mu\delta}f_{G}+
R^{\mu\nu}R^{\gamma}_{\mu\delta \nu}f_{G}+\frac{1}{2}R
\nabla_{\delta}\nabla^{\gamma}f_{G}-\frac{1}{2}R^{\mu\nu n
\gamma}R_{\delta \mu\nu
n}f_{G}\right.\\\nonumber&+&\left.R^{\gamma}_{\delta}\Box f_{G}
-R^{\mu}_{\delta}\nabla_{\mu}\nabla^{\gamma}f_{G}-R^{\mu\gamma}\nabla_{\mu}\nabla_{\delta}f_{G}-R^{\gamma}_{\mu\delta
\nu}\nabla^{\nu}\nabla^{\mu}f_{G}-\frac{1}{2}RR^{\gamma}_{\delta}f_{G}\right]
\\\nonumber &\times&g_{\eta\vartheta}{\mathcal{V}}_{\gamma}{\mathcal{V}}^{\delta}g^{\eta\vartheta}-\left[4R_{\mu m}R^{\mu m}f_{G}
-2R^{2}f_{G}+4R^{m}_{\mu m
\nu}R^{\mu\nu}f_{G}\right.\\\nonumber&-&\left.2R^{l}_{\mu \nu
n}R^{\mu \nu
n}_{l}f_{G}+16R^{\mu\nu}\nabla_{\nu}\nabla_{\mu}f_{G}-4R\Box f_{G}
-4R^{\mu
m}\nabla_{\mu}\nabla_{m}f_{G}\right.\\\nonumber&-&\left.4R^{m}_{\mu
m \nu}\nabla^{\nu}\nabla^{\mu}f_{G}-4R^{\nu
l}\nabla_{\nu}\nabla_{l}f_{G}\right]-6R\Box f_{G}-
\frac{1}{2}f+12R^{\mu\nu}\nabla_{\mu}\nabla_{\nu}f_{G},
\end{eqnarray}
\begin{eqnarray}\nonumber
\textsf{J}^{\textsf{(cor)}}_{(\eta\vartheta)} &=& \left[2R_{\mu
d}R^{\mu}_{c}f_{G}+2R^{\mu\nu}R_{\mu d \nu
c}f_{G}-RR_{cd}f_{G}-R_{d\mu\nu n}R^{\mu\nu n}_{c}f_{G}+2R_{cd}\Box
f_{G}\right.\\\nonumber&+&\left.R\nabla_{d}\nabla_{c}f_{G}-2R^{\mu}_{d}\nabla_{\mu}\nabla_{c}f_{G}-2R^{\mu}_{c}\nabla_{\mu}\nabla_{d}f_{G}
-2R_{\mu d\nu c}\nabla^{\mu}\nabla^{\nu}f_{G}\right]
h^{c}_{\eta}h^{d}_{\vartheta}\\\nonumber&+&2\left[R^{\mu\gamma}R_{\mu\delta}f_{G}+
R^{\gamma}_{\mu\delta
\nu}R^{\mu\nu}f_{G}-\frac{1}{2}RR^{\gamma}_{\delta}f_{G}-\frac{1}{2}R_{\delta
\mu\nu n}R^{\mu\nu n
\gamma}f_{G}\right.\\\nonumber&+&\left.R^{\gamma}_{\delta}\Box
f_{G}-R^{\mu\gamma}\nabla_{\mu}\nabla_{\delta}f_{G}+\frac{1}{2}R
\nabla_{\delta}\nabla^{\gamma}f_{G}
-R^{\mu}_{\delta}\nabla^{\gamma}\nabla_{\mu}f_{G}\right.\\\nonumber&-&\left.R^{\gamma}_{\mu\delta
\nu}\nabla^{\mu}\nabla^{\nu}f_{G}\right]h_{\eta\vartheta}{\mathcal{V}}_{\gamma}{\mathcal{V}}^{\delta}-2\left[R_{\mu\vartheta}R^{\mu}_{\eta}f_{G}+
R^{\mu\nu}R_{\mu\vartheta
\nu\eta}f_{G}-\frac{1}{2}RR_{\eta\vartheta}f_{G}\right.\\\nonumber&-&\left.\frac{1}{2}R_{\vartheta
\mu\nu n}R^{\mu\nu n}_{\eta}f_{G}+R_{\eta\vartheta}\Box
f_{G}+\frac{1}{2}R
\nabla_{\eta}\nabla_{\vartheta}f_{G}-R^{\mu}_{\eta}\nabla_{\vartheta}\nabla_{\mu}f_{G}\right.\\\nonumber&-&\left.
R^{\mu}_{\vartheta}\nabla_{\eta}\nabla_{\mu}f_{G}-R_{\mu\vartheta
\nu\eta}\nabla^{\mu}\nabla^{\nu}f_{G}\right]-2\left[-R_{\mu\delta}R^{\mu}_{\eta}f_{G}-
R^{\mu\nu}R_{\mu\delta
\nu\eta}f_{G}\right.\\\nonumber&+&\left.\frac{1}{2}RR_{\eta\delta}f_{G}+\frac{1}{2}R_{\delta
\mu\nu n}R^{\mu\nu n}_{\eta }f_{G}-R_{\delta\eta}\Box
f_{G}+R_{\mu\delta
\nu\eta}\nabla^{\mu}\nabla^{\nu}f_{G}\right.\\\nonumber&-&\left.\frac{1}{2}R
\nabla_{\eta}\nabla_{\delta}f_{G}+R^{\mu}_{\eta}\nabla_{\delta}\nabla_{\mu}f_{G}
+R^{\mu}_{\delta}\nabla_{\eta}\nabla_{\mu}f_{G}\right]{\mathcal{V}}_{\vartheta}{\mathcal{V}}^{\delta}-2\left[-R_{\mu\vartheta}R^{\mu\gamma}f_{G}\right.
\\\nonumber&-&\left.R^{\mu\nu}R^{\gamma}_{\mu\vartheta
\nu}f_{G}+\frac{1}{2}RR^{\gamma}_{\vartheta}f_{G}+\frac{1}{2}R_{\vartheta
\mu\nu n}R^{\mu\nu n\gamma}f_{G}-R^{\gamma}_{\vartheta}\Box
f_{G}\right.\\\nonumber&+&\left.R^{\mu\gamma}\nabla_{\mu}\nabla_{\vartheta}f_{G}+R^{\mu}_{\vartheta}\nabla_{\mu}\nabla^{\gamma}f_{G}-\frac{1}{2}R
\nabla^{\gamma}\nabla_{\vartheta}f_{G}+R^{\gamma}_{\mu\vartheta
\nu}\nabla^{\mu}\nabla^{\nu}f_{G}\right]{\mathcal{V}}_{\eta}{\mathcal{V}}_{\gamma}\\\nonumber&-&2\left[
R^{\gamma}_{\mu\delta
\nu}R^{\mu\nu}f_{G}+R_{\mu\delta}R^{\mu\gamma}f_{G}-\frac{1}{2}RR^{\gamma}_{\delta}f_{G}-\frac{1}{2}R^{\mu\nu
n \gamma}R_{\delta \mu\nu
n}f_{G}\right.\\\nonumber&+&\left.R^{\gamma}_{\delta}\Box
f_{G}+\frac{1}{2}R
\nabla_{\delta}\nabla^{\gamma}f_{G}-R^{\mu}_{\delta}\nabla_{\mu}\nabla^{\gamma}f_{G}-R^{\mu\gamma}\nabla_{\mu}\nabla_{\delta}f_{G}
\right.\\\nonumber&-&\left.R^{\gamma}_{\mu\delta
\nu}\nabla^{\nu}\nabla^{\mu}f_{G}\right]{\mathcal{V}}_{\gamma}{\mathcal{V}}^{\delta}g_{\eta\vartheta},
\end{eqnarray}
\begin{eqnarray}\nonumber
\textsf{Q}^{\textsf{(cor)}}
&=&\left[\frac{1}{2}R_{\mu\epsilon}R^{\mu
p}f_{G}+\frac{1}{2}R^{\mu\nu}R^{p}_{\mu\epsilon
\nu}f_{G}-\frac{1}{4}RR^{p}_{\epsilon}f_{G}-\frac{1}{4}R_{\epsilon
\mu\nu n}R^{\mu\nu n p}f_{G}\right.\\\nonumber&+&\left.\frac{1}{2}
R^{p}_{\epsilon}\Box
f_{G}+\frac{1}{4}R\nabla^{p}\nabla_{\epsilon}f_{G}-\frac{1}{4}R^{\mu
p}\nabla_{\epsilon}\nabla_{\mu}f_{G}-\frac{1}{2}R^{\mu}_{\epsilon}\nabla^{p}\nabla_{\mu}
f_{G}\right.\\\nonumber&-&\left.\frac{1}{2}R^{p}_{\mu\epsilon
\nu}\nabla^{\mu}\nabla^{\nu}f_{G}\right]g^{\eta\vartheta}\epsilon_{p
\delta\vartheta}\epsilon^{\epsilon\delta}_{\eta}+\left[-\frac{1}{2}R_{\mu\delta}R^{\mu
p}f_{G}-\frac{1}{2}R^{\mu\nu}R^{p}_{\mu\delta \nu}
f_{G}\right.\\\nonumber&+&\left.\frac{1}{4}R^{p}_{\delta}Rf_{G}-\frac{1}{2}R^{p}_{\delta}\Box
f_{G}+\frac{1}{4}R^{\mu\nu np}R_{\delta \mu\nu
n}f_{G}-\frac{1}{4}R\nabla^{p}\nabla_{\delta}f_{G}\right.\\\nonumber&+&\left.\frac{1}{2}R^{\mu}_{\delta}\nabla^{p}\nabla_{\mu}f_{G}+\frac{1}{4}R^{\mu
p}\nabla_{\delta}\nabla_{\mu}f_{G} +\frac{1}{2}R^{p}_{\mu\delta
\nu}\nabla^{\mu}\nabla^{\nu}f_{G}\right]g^{\eta\vartheta}\epsilon_{p\epsilon
\vartheta}\epsilon^{\epsilon\delta}_{\eta}\\\nonumber&+&\left[-\frac{1}{2}R_{\mu\epsilon}R^{\mu\gamma}f_{G}-\frac{1}{2}R^{\mu\nu}R^{\gamma}_{\mu\epsilon
\nu}f_{G}
+\frac{1}{4}RR^{\gamma}_{\epsilon}f_{G}+\frac{1}{4}R_{\epsilon
\mu\nu n}R^{\mu\nu
n\gamma}f_{G}\right.\\\nonumber&-&\left.\frac{1}{2}R^{\gamma}_{\epsilon}\Box
f_{G}+\frac{1}{4}R^{\mu\gamma}\nabla_{\epsilon}\nabla_{\mu}f_{G}-\frac{1}{4}R\nabla^{\gamma}\nabla_{\epsilon}f_{G}
+\frac{1}{2}R^{\mu}_{\epsilon}\nabla_{\mu}\nabla^{\gamma}
f_{G}\right.\\\nonumber&+&\left.\frac{1}{2}R^{\gamma}_{\mu\epsilon
\nu}\nabla^{\nu}\nabla^{\mu}
f_{G}\right]g^{\eta\vartheta}\epsilon_{\delta\gamma\vartheta}\epsilon^{\epsilon\delta}_{\eta}+\left[\frac{1}{2}R^{\mu\nu}R^{\gamma}_{\mu\delta
\nu}f_{G}+\frac{1}{2}R^{\mu\gamma}R_{\mu\delta}f_{G}
\right.\\\nonumber&-&\left.\frac{1}{4}RR^{\gamma}_{\delta}f_{G}+\frac{1}{2}R^{\gamma}_{\delta}\Box
f_{G}-\frac{1}{4}R^{\mu\nu n\gamma}R_{\delta \mu\nu
n}f_{G}+\frac{1}{4}R\nabla_{\delta}\nabla^{\gamma}f_{G}\right.\\\nonumber&-&\left.\frac{1}{4}R^{\mu\gamma}\nabla_{\delta}\nabla_{\mu}f_{G}
-\frac{1}{2}R^{\gamma}_{\mu\delta
\nu}\nabla^{\mu}\nabla^{\nu}f_{G}-\frac{1}{2}R^{\mu}_{\delta}\nabla^{\gamma}\nabla_{\mu}f_{G}\right]
g^{\eta\vartheta}\epsilon_{\epsilon\gamma\vartheta}\epsilon^{\epsilon\delta}_{\eta}\\\nonumber&-&
12R^{\mu\nu}\nabla_{\nu}\nabla_{\mu}f_{G} +6R\Box
f_{G}+\left[4R_{\mu\nu}R^{\mu\nu}f_{G}+\left(\mathcal{U}+\mathcal{P}\right)f_{\mathcal{T}}\right.\\\nonumber&+&\left.4R^{\mu\nu}R^{m}_{\mu\nu
m}f_{G}-2R^{l}_{\mu\nu n}R^{\mu\nu n}_{l}f_{G}-2R^{2}f_{G}-4R\Box
f_{G}\right.\\\nonumber&+&\left.16R^{\mu\nu}\nabla_{\nu}\nabla_{\mu}f_{G}-4R^{\mu
l}\nabla_{l}\nabla_{\mu}f_{G}-4R^{\mu m}\nabla_{\mu}\nabla_{m}f_{G}
\right.\\\nonumber&-&\left.4R^{m}_{\mu m
\nu}\nabla^{\nu}\nabla^{\mu}f_{G}\right]+\frac{f}{2}.
 \end{eqnarray}
\textbf{Data Availability Statement:} This manuscript has no
associated data.

\end{document}